\documentclass[reprint,showpacs,groupedaddress,eqsecnum,amsfonts,amsmath,amssymb,aps,prd]{revtex4}
\usepackage{graphicx}
\usepackage{latexsym}
\usepackage{amssymb}
\usepackage{mathrsfs}
\usepackage{amsmath}
\usepackage{amsthm}
\usepackage{color}
\usepackage{dcolumn}
\usepackage{bm}

\definecolor{dyellow}{rgb}{1.,0.8,.0}
\definecolor{myblue}{rgb}{.1,.1,.7}
\definecolor{dcyan}{rgb}{.0,.6,.6}
\definecolor{dmagenta}{rgb}{0.6,0.0,0.6}
\definecolor{brown}{rgb}{0.6,0.2,0.}
\definecolor{darkblue}{rgb}{.0,.0,0.5}
\definecolor{darkred}{rgb}{0.75,0.0,0.0}
\definecolor{orange}{rgb}{1.,.6,.0}
\definecolor{dorange}{rgb}{0.8,.4,.0}
\definecolor{darkgreen}{rgb}{0.0,0.6,0.0}
\definecolor{purple}{rgb}{.4,.0,.4}
\definecolor{grey}{rgb}{0.5,0.5,0.5}

\begin{document}
\hyphenpenalty=1000
\preprint{APS/123-QED}
\title{Multipole analysis for linearized $f(R,\mathcal{G})$ gravity with irreducible Cartesian tensors}

\newcommand*{\PKU}{Institute of High Energy Physics and Theoretical Physics Center for Science Facilities,
Chinese Academy of Sciences, Beijing, 100049, People's Republic of China}\affiliation{\PKU}
\newcommand*{\INFN}{INFN, Sez. di Pavia, via Bassi 6, 27100 Pavia, Italy}\affiliation{\INFN}
\newcommand*{\CICQM}{}\affiliation{\CICQM}
\newcommand*{\CHEP}{}\affiliation{\CHEP}

\author{Bofeng Wu}\email{wubf@ihep.ac.cn}\affiliation{\PKU}
\author{Chao-Guang Huang}\email{huangcg@ihep.ac.cn}\affiliation{\PKU}

\begin{abstract}
The field equations of $f(R,\mathcal{G})$ gravity are rewritten in the form of obvious wave equations with the stress-energy pseudotensor of the matter fields and the gravitational field, as their sources, under the de Donder condition. The linearized field equations of $f(R,\mathcal{G})$ gravity are the same as those of linearized $f(R)$ gravity, and thus, their multipole expansions under the de Donder condition are also the same. It is also shown that the Gauss-Bonnet curvature scalar $\mathcal{G}$ does not contribute to the effective stress-energy tensor of gravitational waves in linearized $f(R,\mathcal{G})$ gravity, though $\mathcal{G}$ plays an important role in the nonlinear effects in general. Further, by applying the $1/r$ expansion in the distance to the source to the linearized $f(R,\mathcal{G})$ gravity, the energy, momentum, and angular momentum carried by gravitational waves in linearized $f(R,\mathcal{G})$ gravity are provided, which shows that $\mathcal{G}$, unlike the nonlinear term $R^2$ in the gravitational Lagrangian, does not contribute to them either.
\end{abstract}
\maketitle
\section{Introduction}
The detection of gravitational waves (GWs) by the LIGO and Virgo collaborations~\cite{TheLIGOScientific:2016agk}
is a milestone in GWs, and promotes the study of General Relativity (GR) and astrophysics~\cite{TheLIGOScientific:2016src,TheLIGOScientific:2016htt,GBM:2017lvd}.
Their observations of GWs are very consistent with GR's prediction, which further confirms that GR is a great and successful theory of gravity. In despite of this, GR has to face many challenges, e.g., interpreting many data observed at infrared scales~\cite{AstierP2006,Eisenstein2005,Riess2004,Spergel2007}. Introducing the modified gravity theories (MGTs)~\cite{Sotiriou2010} is another approach to deal with these difficulties.

$f(R)$ gravity~\cite{Buchdahl1970,Starobinsky:1980te,York1972,Gibbons1977,Nojiri:2010wj,DNojiri:2017ncd} is one of the
simplest MGTs, and it replaces the Einstein-Hilbert action by the quantity $f(R)$ in the gravitational Lagrangian, where $f$ is a general function of the Ricci scalar $R$. Another typical MGT is $f(R,\mathcal{G})$ gravity~\cite{Cognola:2006eg,Alimohammadi:2008fq,Wu:2015maa,Shamir:2017ndy}, which further generalizes GR by adopting a general function of $R$ and the Gauss-Bonnet (GB) curvature scalar $\mathcal{G}$ in the gravitational Lagrangian.
In this paper, we only consider polynomial $f(R,\mathcal{G})$ models of the form
\begin{align}
\label{equ1.1}&f(R,\mathcal{G})=R+a_{2}\mathcal{G}+a_{11}R^{2}+a_{12}R\mathcal{G}+a_{22}\mathcal{G}^2+a_{111}R^3
+a_{112}R^2\mathcal{G}+a_{122}R\mathcal{G}^2+a_{222}\mathcal{G}^3+\cdots,
\end{align}
where $a_{2},a_{11}\cdots$ are the coupling constants. Because the integral of $\mathcal{G}$ over a 4-dimensional manifold is a topological invariant, the term $a_{2}\mathcal{G}$ in the above Lagrangian does not contribute to the field equation of $f(R,\mathcal{G})$ gravity. Under the post-Minkowskian method, the coupling constant $a_{12}$ will appear in the second-order field equation of $f(R,\mathcal{G})$ gravity. This implies that $\mathcal{G}$ will play an important role in the nonlinear effects, and therefore, it is worthwhile to investigate the nonlinear effects contributed by $\mathcal{G}$.

The multipole analysis is the most useful way to describe the external field for the sources localized in a finite region of space~\cite{Damour:1990gj}. The symmetric and trace-free (STF) formalism in terms of the irreducible Cartesian tensors, developed by Thorne~\cite{Thorne:1980ru} and Blanchet and Damour~\cite{Blanchet:1985sp,Blanchet:1989ki}, is one of the important methods with respect to the multipole analysis, and the relevant STF technique is summarized in Ref.~\cite{Damour:1990gj}. The multipole analysis in terms of the STF formalism can be used to explore the nonlinear effects of a gravitational theory~\cite{Thorne:1980ru}. In this paper, we will apply the STF formalism to $f(R,\mathcal{G})$ gravity to investigate its related nonlinear effects.

In our preceding papers~\cite{Wu:2017vvm,Wu:2018hjx}, referred to as I and II hereafter, respectively,
the multipole analysis for linearized $f(R)$ gravity with irreducible Cartesian tensors has been presented, and the relevant results include:
\begin{itemize}
\item The field equations of $f(R)$ gravity are rewritten in the form of obvious wave equations with the stress-energy pseudotensor of the matter fields and the gravitational field, as the sources, under de Donder condition. This process is similar to that of GR~\cite{Thorne:1980ru,Blanchet:2013haa};
\item The field equations and the effective stress-energy tensor of GWs for linearized $f(R)$ gravity are presented, and the latter is a typical second-order nonlinear quantity;
\item The STF formalism is applied to the linearized $f(R)$ gravity, and its explicit expression of multipole expansion is derived;
\item The $1/r$ expansion in the distance to the source is applied to the linearized $f(R)$ gravity, and its multipole expansion in the radiation field with irreducible Cartesian tensors is presented;
\item As the further second-order nonlinear quantities, the energy, momentum, and angular momentum carried by GWs in linearized $f(R)$ gravity are provided.
\end{itemize}
In this paper, we will follow the above process to discuss the multipole analysis for linearized $f(R,\mathcal{G})$ gravity with irreducible Cartesian tensors.

For $f(R,\mathcal{G})$ gravity, similarly to $f(R)$ gravity~\cite{Wu:2017vvm}, one has to introduce
the gravitational field amplitude
\begin{equation}\label{equ1.2}
h^{\mu\nu}:=\sqrt{-g}g^{\mu\nu}-\eta^{\mu\nu}
\end{equation}
and the effective gravitational field amplitude
\begin{equation}\label{equ1.3}
\tilde{h}^{\mu\nu}:=f_{R}\sqrt{-g}g^{\mu\nu}-\eta^{\mu\nu},
\end{equation}
where $g^{\mu\nu}$ denotes the contravariant metric, $\eta^{\mu\nu}$ represents an auxiliary Minkowskian metric, $g$ is the determinant of metric $g_{\mu\nu}$, and $f_{R}=\partial_{R}f$. With the help of $h^{\mu\nu}$ and $\tilde{h}^{\mu\nu}$, by using the same method in Refs.~\cite{fockv,Wu:2017vvm}, the field
equations of $f(R,\mathcal{G})$ gravity can be rewritten in the form of obvious wave equations under the de Donder condition, and the source terms, composed of all the nonlinear terms under the post-Minkowskian method, are the stress-energy pseudotensor of the matter fields and the gravitational field. If $\tilde{h}^{\mu\nu}$  is a perturbation, the resulting field equations of linearized $f(R,\mathcal{G})$ gravity are the same as those of linearized $f(R)$ gravity~\cite{Wu:2017vvm}, as expected. Furthermore, under this condition, the effective stress-energy tensor of GWs in linearized $f(R,\mathcal{G})$ gravity, as a typical second-order nonlinear quantity, is shown to be the same as that of linearized $f(R)$ gravity, which implies that
the GB scalar $\mathcal{G}$ does not contribute to the effective stress-energy tensor of GWs in linearized $f(R,\mathcal{G})$ gravity, though $\mathcal{G}$ usually plays an important role in the nonlinear effects as mentioned before.

As the further second-order nonlinear quantities, the energy, momentum, and angular momentum carried by GWs in linearized $f(R,\mathcal{G})$ gravity need to be discussed. Similarly to GR~\cite{Thorne:1980ru} and $f(R)$ gravity~\cite{Wu:2018hjx}, the energy and the momentum can be derived directly by use of the effective stress-energy tensor of GWs, but the angular momentum cannot. So, a new way, not depending on the effective stress-energy tensor of GWs, need to be found to deal with the angular momentum. In this paper, by following Refs.~\cite{Thorne:1980ru,Peters:1964zz,Wu:2018hjx}, the energy, momentum, and angular momentum carried by GWs in linearized $f(R,\mathcal{G})$ gravity will be dealt with in a unified way.

This unified way requires that the multipole expansion of the linearized $f(R,\mathcal{G})$ gravity is discussed firstly. By definitions (\ref{equ1.2}) and (\ref{equ1.3}), the true gravitational field amplitude $h^{\mu\nu}$ in $f(R,\mathcal{G})$ gravity can be read out from $\tilde{h}^{\mu\nu}$.  For the linearized $f(R,\mathcal{G})$ gravity, the relation between $h^{\mu\nu}$ and $\tilde{h}^{\mu\nu}$ becomes linear
and simple, namely,
\begin{equation}
\label{equ1.4}h^{\mu\nu}=\tilde{h}^{\mu\nu}-2a_{11}R^{(1)}\eta^{\mu\nu},
\end{equation}
where $R^{(1)}$ is the linear part of Ricci scalar $R$. This relation is the same as that in $f(R)$ gravity. As mentioned above,
the linearized $f(R,\mathcal{G})$ gravity and the linearized $f(R)$ gravity have the same field equations, so
the multipole expansions of $\tilde{h}^{\mu\nu}$ under the de Donder condition in these two gravitational models
are also the same. Moreover, for these two models, $R^{(1)}$
satisfies the same Klein-Gordon (KG) equation with an external source and has the same multipole expansion.
Therefore, Eq.~\eqref{equ1.4} implies that these two gravitational models have the same
multipole expansion, and in particular, they have the same multipole expansion in the radiation field. 

With the help of the relevant STF technique, the $1/r$ expansion in the distance to the source can be applied to the linearized $f(R,\mathcal{G})$ gravity. Based on this, the energy, momentum, and angular momentum carried by GWs in linearized $f(R,\mathcal{G})$ gravity can be dealt with in the above unified way. Thus, we present the explicit expressions of the total power and rates of momentum and angular momentum carried by GWs in linearized $f(R,\mathcal{G})$ gravity, and prove that the terms associated with GB curvature scalar $\mathcal{G}$ in the above three quantities are exactly cancelled each other after the average over a small spatial volume.  In one sentence, the GB curvature scalar $\mathcal{G}$ does not contribute to the total power and rates of momentum and angular moment carried by GWs in linearized $f(R,\mathcal{G})$ gravity.

This paper is organized as follows. In Sec.~\ref{Sec:Preliminary}, the notation, the
relevant formulas of STF formalism, and the review of the metric $f(R,\mathcal{G})$ gravity are described.
In Sec.~\ref{Sec:Equation}, by using the same method in Refs.~\cite{fockv,Wu:2017vvm}, we show that the
field equations of $f(R,\mathcal{G})$ gravity can be rewritten in the form of obvious wave equations with the stress-energy pseudotensor of the matter fields and the gravitational field, as the sources, under de Donder condition. Upon this, we derive the field equations and the effective stress-energy tensor of GWs for linearized $f(R,\mathcal{G})$ gravity. In Sec.~\ref{Sec:EneggyMomentumAgular}, we show that the linearized
$f(R,\mathcal{G})$ gravity and the linearized $f(R)$ gravity have the same multipole expansion. Then, we present its expression, and derive the corresponding expression in the radiation field. Further, by using the $1/r$ expansion in the distance to the source. we evaluate the energy, momentum, and angular momentum carried by GWs in linearized $f(R,\mathcal{G})$ gravity systematically. In Sec.~\ref{Sec:Conclusion}, we present the conclusions and make some discussions.
\section{Preliminary\label{Sec:Preliminary}}
\subsection{Notation~\label{Sec:Notation}}
The notation in this paper is the same as that in I and II~\cite{Wu:2017vvm,Wu:2018hjx}. The international system of units is used throughout the paper. The signature of the metric $g^{\mu\nu}$ is $(-,+,+,+)$. $\epsilon_{ijk}$ is the three-dimensional Levi-Civit\`{a} symbol with $\epsilon_{123}=1$. The greek indices run from 0 to 3 and the latin indices run from 1 to 3, where repeated indices indicate to be summed. In the linearized gravitational theory, $(x^{\mu})=(ct,x_{i})$ behave as though they were Minkowskian coordinates. The spherical coordinate system $(ct,r,\theta,\varphi)$ is defined by
\begin{equation}\label{equ2.1}
x_{1}=r\sin{\theta}\cos{\varphi},\ x_{2}=r\sin{\theta}\sin{\varphi},\ x_{3}=r\cos{\theta}.
\end{equation}
As in the flat space, $\boldsymbol{x}$ is used to denote the radial vector, and $x_{i}$ in Eq.~\eqref{equ2.1} are its components. $\boldsymbol{n}=\boldsymbol{x}/r=(n_{i})$ denotes the unit radial vector, where $r$ is the length of $\boldsymbol{x}$, and $n_{i}=x_{i}/r$ are the components of $\boldsymbol{n}$. Obviously, by Eq.~(\ref{equ2.1}),
\begin{equation}\label{equ2.2}
\partial_{r}=\frac{\partial}{\partial r}=n_{1}\frac{\partial}{\partial x_{1}}+n_{2}\frac{\partial}{\partial x_{2}}+n_{3}\frac{\partial}{\partial x_{3}}=n_{i}\partial_{i}.
\end{equation}

The symbol
\begin{equation}\label{equ2.3}
B_{I_{l}}\equiv B_{i_{1}i_{2}\cdots i_{l}}
\end{equation}
is used to denote the Cartesian tensor with $l$ indices~\cite{Thorne:1980ru}, and especially, the tensor products of $l$ radial and unit radial vectors are abbreviated by
\begin{align}
\label{equ2.4}X_{I_{l}}=X_{i_{1}i_{2}\cdots i_{l}}:= x_{i_{1}}x_{i_{2}}\cdots x_{i_{l}},\\
\label{equ2.5}N_{I_{l}}=N_{i_{1}i_{2}\cdots i_{l}}:= n_{i_{1}}n_{i_{2}}\cdots n_{i_{l}}
\end{align}
with
\begin{equation}\label{equ2.6}
X_{I_{l}}=r^l N_{I_{l}}.
\end{equation}

\subsection{Relevant formulas in STF formalism~\label{Sec:STFformulae}}
In this subsection, the relevant formulas in the STF formalism are listed~\cite{Damour:1990gj,Thorne:1980ru,Blanchet:1985sp,Wu:2017vvm,Wu:2018hjx}.
The symmetric part and the STF part of a Cartesian tensor $B_{I_{l}}$ are expressed by
\begin{align}
\label{equ2.7}B_{(I_{l})}&=B_{(i_{1}i_{2}\cdots i_{l})}:=\frac{1}{l!}\sum_{\sigma} B_{i_{\sigma(1)}i_{\sigma(2)}\cdots i_{\sigma(l)}},\\
\label{equ2.8}\hat{B}_{I_{l}}&\equiv B_{<I_{l}>}\equiv B_{<i_{1}i_{2}\cdots i_{l}>}
:=\sum_{k=0}^{[\frac{l}{2}]}b_{k}\delta_{(i_{1}i_{2}}\cdots\delta_{i_{2k-1}i_{2k}}S_{i_{2k+1}\cdots i_{l})a_{1}a_{1}\cdots a_{k}a_{k}},
\end{align}
respectively, where $\sigma$ runs over all permutations of $(12\cdots l)$,
\begin{align}
\label{equ2.9}S_{I_{l}}=&B_{(I_{l})},\\
\label{equ2.10}b_{k}=&(-1)^{k}\frac{(2l-2k-1)!!}{(2l-1)!!}\frac{l!}{(2k)!!(l-2k)!}.
\end{align}
Moreover, there are
\begin{align}
\label{equ2.11}\hat{N}_{I_{l}}&=\sum_{k=0}^{[\frac{l}{2}]}b_{k}\delta_{(i_{1}i_{2}}\cdots\delta_{i_{2k-1}i_{2k}}
N_{i_{2k+1}\cdots i_{l})},\\
\label{equ2.12}\hat{\partial}_{I_{l}}&=\sum_{k=0}^{[\frac{l}{2}]}b_{k}\delta_{(i_{1}i_{2}}\cdots\delta_{i_{2k-1}i_{2k}}
\partial_{i_{2k+1}\cdots i_{l})},\\
\label{equ2.13}\partial_{i}n_{j}&=\frac{1}{r}(\delta_{ij}-n_{i}n_{j}),\\
\label{equ2.14}\partial_{i}\left(F\left(t-\frac{\epsilon r}{c}\right)\right)&=-\frac{\epsilon n_{i}}{c}\partial_{t}F\left(t-\frac{\epsilon r}{c}\right),\qquad (\epsilon^{2}=1),
\end{align}
\begin{align}
\label{equ2.15}\hat{\partial}_{I_{l}}\left(\frac{F(t-\epsilon r/c)}{r}\right)&=(-\epsilon)^{l}\hat{N}_{I_{l}}\sum_{k=0}^{l}\frac{(l+k)!}{(2\epsilon)^{k}k!(l-k)!}
\frac{\partial_{t}^{l-k}F(t-\epsilon r/c)}{c^{l-k}r^{k+1}},\qquad (\epsilon^{2}=1),
\end{align}
where $\partial_{I_{l}}\equiv\partial_{i_{1}i_{2}\cdots i_{l}}:=\partial_{i_{1}}\partial_{i_{2}}\cdots\partial_{i_{l}}$
and $\partial_t^{l-k}$ is the $(l-k){\rm th}$ derivative with respect to $t$.

The integral formulas over a unit sphere are
\begin{align}
\label{equ2.16}&\int N_{I_{2l+1}}d\Omega=0,\\
\label{equ2.17}&\int N_{I_{2l}}d\Omega=\frac{4\pi}{2l+1}\delta_{(i_{1}i_{2}}\cdots\delta_{i_{2l-1}i_{2l})},
\end{align}
where $l$ is the non-negative integer and $d\Omega$ is the element of the solid angle about the radial vector.
\subsection{Review of the metric $f(R,\mathcal{G})$ gravity~\label{Sec:f(R)gravity}}
We restrict our attention to the metric $f(R,\mathcal{G})$ gravity~\cite{Wu:2015maa}, and its action is
\begin{equation}\label{equ2.20}
S=\frac{1}{2\kappa c}\int dx^4\sqrt{-g}f(R,\mathcal{G})+S_{M}(g^{\mu\nu},\psi),
\end{equation}
where $f$ is an arbitrary function of the Ricci scalar $R$ and the GB curvature scalar $\mathcal{G}$, $\kappa=8\pi G/c^{4}$, and $S_{M}(g^{\mu\nu},\psi)$ is the matter action. Varying this action with respect to the metric $g^{\mu\nu}$ gives the gravitational field equations and the corresponding trace equation
\begin{equation}\label{equ2.21}
H_{\mu\nu}=\kappa T_{\mu\nu},\qquad H=\kappa T,
\end{equation}
where
\begin{align}
\label{equ2.22}H_{\mu\nu}=&-\frac{g_{\mu\nu}}{2}f+R_{\mu\nu}f_{R}+g_{\mu\nu}\square f_{R}-\nabla_{\mu}\nabla_{\nu} f_{R}+\frac{g_{\mu\nu}}{2}f_{\mathcal{G}}\mathcal{G}+2R g_{\mu\nu}\square f_{\mathcal{G}}-2R\nabla_{\mu}\nabla_{\nu}f_{\mathcal{G}}\notag\\
&+4R_{\nu}^{\phantom{\nu}\lambda}\nabla_{\lambda}\nabla_{\mu}f_{\mathcal{G}}+
4R_{\mu}^{\phantom{\mu}\lambda}\nabla_{\lambda}\nabla_{\nu}f_{\mathcal{G}}-4g_{\mu\nu}R^{\alpha\beta}
\nabla_{\alpha}\nabla_{\beta}f_{\mathcal{G}}-4R_{\mu\nu}\square f_{\mathcal{G}}+4R_{\mu\rho\nu\sigma}\nabla^{\rho}\nabla^{\sigma}f_{\mathcal{G}},\\
\label{equ2.23}H=&-2f+R f_{R}+3\square f_{R}+2f_{\mathcal{G}}\mathcal{G}+2R\square f_{\mathcal{G}}-4R^{\alpha\beta}\nabla_{\alpha}\nabla_{\beta}f_{\mathcal{G}}
\end{align}
with $f_{\mathcal{G}}=\partial_{\mathcal{G}}f$,
and $T_{\mu\nu}$
is the stress-energy tensor of matter fields. In the present paper, only the polynomial $f(R,\mathcal{G})$ models, namely Eq.~\eqref{equ1.1}, is considered.

\section{The field equations and stress-energy pseudotensor of $f(R,\mathcal{G})$ gravity under de Donder condition\label{Sec:Equation}}
\subsection{Obvious wave equation in $f(R,\mathcal{G})$ gravity}
By following GR~\cite{fockv,Blanchet:2013haa} and $f(R)$ gravity~\cite{Wu:2017vvm}, we begin to rewrite the field equations of $f(R,\mathcal{G})$ gravity in the form of obvious wave equations in a fictitious flat spacetime
under the de Donder condition. Firstly, by Eq.~\eqref{equ1.2}, the gravitational field amplitude $h^{\mu\nu}$ is defined as
\begin{align}
\label{equ3.1}h^{\mu\nu}&:=\overline{g}^{\mu\nu}-\eta^{\mu\nu},
\end{align}
where
\begin{align}
\label{equ3.2}\overline{g}^{\mu\nu}&:=\sqrt{-g}g^{\mu\nu}
\end{align}
is the densitized inverse metric. It should be noted that $h^{\mu\nu}$ in Eq.~\eqref{equ3.1} is not necessarily a perturbation. According to Ref.~\cite{fockv},
\begin{align}
\label{equ3.4}\Gamma^{\lambda\mu\nu}:&=g^{\mu\alpha}g^{\nu\beta}\Gamma^{\lambda}_{\alpha\beta}=\frac{1}{2g}(\overline{g}^{\mu\rho}\partial_{\rho}\overline{g}^{\nu\lambda}
+\overline{g}^{\nu\rho}\partial_{\rho}\overline{g}^{\mu\lambda}-\overline{g}^{\lambda\rho}\partial_{\rho}\overline{g}^{\mu\nu})
+\frac{1}{2}(y^{\mu}g^{\nu\lambda}+y^{\nu}g^{\mu\lambda}-y^{\lambda}g^{\mu\nu}),\\
\label{equ3.5}y^{\mu}:&=g^{\mu\rho}\partial_{\rho}\ln{\sqrt{-g}}.
\end{align}
In terms of $h^{\mu\nu}$, the Christoffel symbols read
\begin{align}
\label{equ3.6}\Gamma^{\mu}_{\sigma\nu}=&-\frac{1}{2\sqrt{-g}}(g_{\nu\alpha}\partial_{\sigma}h^{\alpha\mu}
+g_{\sigma\alpha}\partial_{\nu}h^{\alpha\mu}-g_{\sigma\alpha}g_{\nu\beta}g^{\mu\rho}\partial_{\rho}h^{\alpha\beta})+\frac{1}{4\sqrt{-g}}(\delta^{\mu}_{\nu}g_{\epsilon\pi}\partial_{\sigma}h^{\epsilon\pi}
+\delta^{\mu}_{\sigma}g_{\epsilon\pi}\partial_{\nu}h^{\epsilon\pi}-g_{\sigma\nu}g_{\epsilon\pi}g^{\mu\rho}\partial_{\rho}h^{\epsilon\pi}),
\end{align}
and then, the Riemann tensor, Ricci tensor, and Ricci scalar read, respectively,
\begin{align}
\label{equ3.7}R^{\mu\nu\rho\sigma}&=-\frac{1}{2\sqrt{-g}}\bigg(
g^{\rho\epsilon}g^{\nu\pi}\partial_{\epsilon}\partial_{\pi}h^{\mu\sigma}
-g^{\sigma\epsilon}g^{\nu\pi}\partial_{\epsilon}\partial_{\pi}h^{\mu\rho}
+g^{\sigma\epsilon}g^{\mu\pi}\partial_{\epsilon}\partial_{\pi}h^{\nu\rho}
-g^{\rho\epsilon}g^{\mu\pi}\partial_{\epsilon}\partial_{\pi}h^{\nu\sigma}
+\frac{1}{2}\big(g^{\mu\rho}g^{\sigma\epsilon}g^{\nu\pi}
-g^{\mu\sigma}g^{\rho\epsilon}g^{\nu\pi}\notag\\
&+g^{\nu\sigma}g^{\rho\epsilon}g^{\mu\pi}
-g^{\nu\rho}g^{\sigma\epsilon}g^{\mu\pi}\big)g_{\alpha\beta}\partial_{\epsilon}\partial_{\pi}h^{\alpha\beta}\bigg)
-\frac{1}{2g}\bigg(\frac{1}{2}\big(g^{\rho\delta}\partial_{\delta}h^{\mu\lambda}\partial_{\lambda}h^{\nu\sigma}
-g^{\sigma\delta}\partial_{\delta}h^{\mu\lambda}\partial_{\lambda}h^{\nu\rho}
+g^{\sigma\delta}\partial_{\delta}h^{\nu\lambda}\partial_{\lambda}h^{\mu\rho}\notag\\
&-g^{\rho\delta}\partial_{\delta}h^{\nu\lambda}\partial_{\lambda}h^{\mu\sigma}
+g^{\mu\delta}\partial_{\delta}h^{\rho\lambda}\partial_{\lambda}h^{\nu\sigma}
-g^{\mu\delta}\partial_{\delta}h^{\sigma\lambda}\partial_{\lambda}h^{\nu\rho}
+g^{\nu\delta}\partial_{\delta}h^{\sigma\lambda}\partial_{\lambda}h^{\mu\rho}
-g^{\nu\delta}\partial_{\delta}h^{\rho\lambda}\partial_{\lambda}h^{\mu\sigma}
+g^{\lambda\tau}\partial_{\lambda}h^{\mu\sigma}\partial_{\tau}h^{\nu\rho}\notag\\
&-g^{\lambda\tau}\partial_{\lambda}h^{\mu\rho}\partial_{\tau}h^{\nu\sigma}\big)
+\frac{1}{2}g_{\alpha\beta}\big(g^{\rho\epsilon}g^{\sigma\pi}\partial_{\epsilon}h^{\nu\beta}\partial_{\pi}h^{\mu\alpha}
-g^{\sigma\epsilon}g^{\rho\pi}\partial_{\epsilon}h^{\nu\beta}\partial_{\pi}h^{\mu\alpha}
+g^{\mu\epsilon}g^{\nu\pi}\partial_{\epsilon}h^{\sigma\beta}\partial_{\pi}h^{\rho\alpha}
-g^{\mu\epsilon}g^{\nu\pi}\partial_{\epsilon}h^{\rho\beta}\partial_{\pi}h^{\sigma\alpha}\notag\\
&+g^{\rho\epsilon}g^{\nu\pi}\partial_{\epsilon}h^{\mu\alpha}\partial_{\pi}h^{\sigma\beta}
-g^{\sigma\epsilon}g^{\nu\pi}\partial_{\epsilon}h^{\mu\alpha}\partial_{\pi}h^{\rho\beta}
+g^{\sigma\epsilon}g^{\mu\pi}\partial_{\epsilon}h^{\nu\alpha}\partial_{\pi}h^{\rho\beta}
-g^{\rho\epsilon}g^{\mu\pi}\partial_{\epsilon}h^{\nu\alpha}\partial_{\pi}h^{\sigma\beta}
+2g^{\rho\epsilon}g^{\nu\pi}\partial_{\epsilon}h^{\sigma\beta}\partial_{\pi}h^{\mu\alpha}\notag\\
&-2g^{\sigma\epsilon}g^{\nu\pi}\partial_{\epsilon}h^{\rho\beta}\partial_{\pi}h^{\mu\alpha}
+2g^{\sigma\epsilon}g^{\mu\pi}\partial_{\epsilon}h^{\rho\beta}\partial_{\pi}h^{\nu\alpha}
-2g^{\rho\epsilon}g^{\mu\pi}\partial_{\epsilon}h^{\sigma\beta}\partial_{\pi}h^{\nu\alpha}\big)
+\frac{1}{4}\big(g^{\sigma\delta}g^{\rho\nu}\partial_{\delta}h^{\mu\lambda}
-g^{\rho\delta}g^{\sigma\nu}\partial_{\delta}h^{\mu\lambda}\notag\\
&+g^{\rho\delta}g^{\sigma\mu}\partial_{\delta}h^{\nu\lambda}
-g^{\sigma\delta}g^{\rho\mu}\partial_{\delta}h^{\nu\lambda}
+g^{\mu\delta}g^{\rho\nu}\partial_{\delta}h^{\sigma\lambda}
-g^{\mu\delta}g^{\sigma\nu}\partial_{\delta}h^{\rho\lambda}
+g^{\nu\delta}g^{\sigma\mu}\partial_{\delta}h^{\rho\lambda}
-g^{\nu\delta}g^{\rho\mu}\partial_{\delta}h^{\sigma\lambda}\big)g_{\alpha\beta}\partial_{\lambda}h^{\alpha\beta}\notag\\
&+\frac{1}{4}\big(g^{\sigma\nu}\partial_{\lambda}h^{\rho\mu}-g^{\rho\nu}\partial_{\lambda}h^{\sigma\mu}
+g^{\rho\mu}\partial_{\lambda}h^{\sigma\nu}-g^{\sigma\mu}\partial_{\lambda}h^{\rho\nu}\big)g^{\lambda\tau}g_{\alpha\beta}\partial_{\tau}h^{\alpha\beta}
+\frac{1}{8}g_{\alpha\beta}g_{\lambda\tau}\big(g^{\rho\epsilon}g^{\mu\pi}g^{\sigma\nu}-g^{\sigma\epsilon}g^{\mu\pi}g^{\rho\nu}\notag\\
&+g^{\sigma\epsilon}g^{\nu\pi}g^{\rho\mu}-g^{\rho\epsilon}g^{\nu\pi}g^{\sigma\mu}\big)\big(4\partial_{\epsilon}h^{\beta\tau}\partial_{\pi}h^{\alpha\lambda}
+\partial_{\epsilon}h^{\lambda\tau}\partial_{\pi}h^{\alpha\beta}\big)
+\frac{1}{8}\big(g^{\mu\sigma}g^{\rho\nu}-g^{\mu\rho}g^{\sigma\nu}\big)g^{\lambda\tau}g_{\alpha\beta}g_{\epsilon\pi}\partial_{\lambda}h^{\alpha\beta}\partial_{\tau}h^{\epsilon\pi}
\bigg),\\
R^{\mu\nu}&=\frac{1}{2\sqrt{-g}}g^{\alpha\beta}\partial_{\alpha}\partial_{\beta}h^{\mu\nu}-\frac{1}{2g}g^{\mu\alpha}g_{\beta\tau}\partial_{\lambda}h^{\nu\tau}\partial_{\alpha}h^{\beta\lambda}+
\frac{1}{4g}g^{\mu\alpha}g^{\nu\beta}g_{\lambda\tau}g_{\epsilon\pi}\partial_{\alpha}h^{\lambda\pi}\partial_{\beta}h^{\tau\epsilon}+\frac{1}{2g}g_{\alpha\beta}g^{\lambda\tau}\partial_{\lambda}h^{\mu\alpha}\partial_{\tau}h^{\nu\beta}\notag\\
&+\frac{1}{2g}\partial_{\alpha}h^{\mu\beta}\partial_{\beta}h^{\nu\alpha}-\frac{1}{2g}g^{\nu\alpha}g_{\beta\tau}\partial_{\lambda}h^{\mu\tau}\partial_{\alpha}h^{\beta\lambda}-\frac{1}{8g}g^{\mu\alpha}g^{\nu\beta}g_{\tau\epsilon}g_{\lambda\pi}\partial_{\alpha}h^{\lambda\pi}\partial_{\beta}h^{\tau\epsilon}
-\frac{1}{4g}g^{\mu\nu}g_{\rho\tau}g_{\epsilon\sigma}g^{\alpha\beta}\partial_{\alpha}h^{\rho\sigma}\partial_{\beta}h^{\tau\epsilon}\notag\\
&-\frac{1}{4\sqrt{-g}}g^{\mu\nu}g^{\alpha\beta}g_{\rho\sigma}\partial_{\alpha}\partial_{\beta}h^{\rho\sigma}-\frac{1}{2\sqrt{-g}}g^{\mu\alpha}\partial_{\alpha}\partial_{\lambda}h^{\nu\lambda}-\frac{1}{2\sqrt{-g}}g^{\nu\alpha}\partial_{\alpha}\partial_{\lambda}h^{\mu\lambda}
+\frac{1}{4g}g^{\mu\nu}g_{\rho\sigma}\partial_{\alpha}h^{\rho\sigma}\partial_{\lambda}h^{\alpha\lambda}\notag\\
\label{equ3.8}&-\frac{1}{2g}\partial_{\alpha}h^{\mu\nu}\partial_{\lambda}h^{\alpha\lambda},\\
R&=-\frac{1}{4g}g^{\alpha\beta}g_{\rho\tau}g_{\epsilon\sigma}\partial_{\alpha}h^{\rho\sigma}\partial_{\beta}h^{\tau\epsilon}-\frac{1}{2\sqrt{-g}}g_{\rho\sigma}g^{\alpha\beta}\partial_{\alpha}\partial_{\beta}h^{\rho\sigma}
-\frac{1}{\sqrt{-g}}\partial_{\alpha}\partial_{\beta}h^{\alpha\beta}
+\frac{1}{2g}g_{\rho\sigma}\partial_{\alpha}h^{\rho\sigma}\partial_{\lambda}h^{\alpha\lambda}\notag\\
\label{equ3.9}&-\frac{1}{2g}g_{\mu\nu}\partial_{\lambda}h^{\mu\tau}\partial_{\tau}h^{\nu\lambda}
-\frac{1}{8g}g^{\mu\nu}g_{\tau\epsilon}g_{\lambda\pi}\partial_{\mu}h^{\lambda\pi}\partial_{\nu}h^{\tau\epsilon}.
\end{align}

In order to apply the de Donder condition, we need to define the effective gravitational field amplitude $\tilde{h}^{\mu\nu}$ by Eq.~\eqref{equ1.3}, namely,
\begin{align}
\label{equ3.10}\tilde{h}^{\mu\nu}&:=\tilde{g}^{\mu\nu}-\eta^{\mu\nu},\\
\label{equ3.11}\tilde{g}^{\mu\nu}&:=f_{R}\sqrt{-g}g^{\mu\nu}.
\end{align}
Obviously, besides the information of metric, it also contains the information of the function $f_{R}$,
which is, from (\ref{equ1.1}),
\begin{equation}\label{equ3.12}
f_{R}=1+2a_{11}R+a_{12}\mathcal{G}+3a_{111}R^{2}+2a_{112}R\mathcal{G}+a_{122}\mathcal{G}^2+\cdots.
\end{equation}
For GR, the de Donder condition is the condition for the harmonic coordinates~\cite{fockv,Blanchet:2013haa}:
\begin{equation}\label{equ3.13}
\square x^\mu =0\Leftrightarrow \partial_{\mu}\overline{g}^{\mu\nu}=\partial_{\mu}h^{\mu\nu}=0.
\end{equation}
For $f(R)$ gravity~\cite{Wu:2017vvm}, the de Donder condition in GR has been generalized to
\begin{equation}\label{equ3.14}
\partial_{\mu}\tilde{g}^{\mu\nu}=\partial_{\mu}\tilde{h}^{\mu\nu}=0\Leftrightarrow \square x^\mu =-g^{\mu\nu}\partial_\nu\ln f_R,
\end{equation}
where the definitions of $\tilde{h}^{\mu\nu}$ and $\tilde{g}^{\mu\nu}$ are the same as Eqs.~\eqref{equ3.10}
and \eqref{equ3.11}, respectively. In this paper, we also adopt the de Donder condition \eqref{equ3.14} for
$f(R,\mathcal{G})$ gravity.

Now, we will express the Riemann tensor, the Ricci tensor, and
the Ricci scalar in terms of $\tilde h^{\mu\nu}$ with the help of the de Donder condition (\ref{equ3.14}).
By Eqs.~(\ref{equ3.1}), (\ref{equ3.2}), (\ref{equ3.10}), and (\ref{equ3.11}), there is
\begin{align}
\label{equ3.15}h^{\mu\nu}&=\frac{1}{f_{R}}\tilde{h}^{\mu\nu}+\left(\frac{1}{f_{R}}-1\right)\eta^{\mu\nu}.
\end{align}
It immediately results in
\begin{align}
\label{equ3.16}\partial_{\lambda}h^{\mu\nu}&=\frac{1}{f_{R}}\partial_{\lambda}\tilde{h}^{\mu\nu}
-\overline{g}^{\mu\nu}\partial_{\lambda}\ln{f_{R}}.
\end{align}
With the help of the de Donder condition (\ref{equ3.14}), the substitution of (\ref{equ3.15}) and (\ref{equ3.16}) in
Eqs.~(\ref{equ3.7}), (\ref{equ3.8}) and \eqref{equ3.9} gives rise to the expressions of $R^{\mu\nu\rho\sigma}$, $R^{\mu\nu}$ and $R$ in terms of $\tilde h^{\mu\nu}$:
\begin{align}
\label{equ3.17}R^{\mu\nu\rho\sigma}&=-\frac{1}{2f_{R}\sqrt{-g}}\bigg(
g^{\rho\epsilon}g^{\nu\pi}\partial_{\epsilon}\partial_{\pi}\tilde{h}^{\mu\sigma}
-g^{\sigma\epsilon}g^{\nu\pi}\partial_{\epsilon}\partial_{\pi}\tilde{h}^{\mu\rho}
+g^{\sigma\epsilon}g^{\mu\pi}\partial_{\epsilon}\partial_{\pi}\tilde{h}^{\nu\rho}
-g^{\rho\epsilon}g^{\mu\pi}\partial_{\epsilon}\partial_{\pi}\tilde{h}^{\nu\sigma}
+\frac{1}{2}\big(g^{\mu\rho}g^{\sigma\epsilon}g^{\nu\pi}
-g^{\mu\sigma}g^{\rho\epsilon}g^{\nu\pi}\notag\\
&+g^{\nu\sigma}g^{\rho\epsilon}g^{\mu\pi}
-g^{\nu\rho}g^{\sigma\epsilon}g^{\mu\pi}\big)g_{\alpha\beta}\partial_{\epsilon}\partial_{\pi}\tilde{h}^{\alpha\beta}\bigg)
-\frac{1}{2gf_{R}^{2}}\bigg(\frac{1}{2}\big(g^{\rho\delta}\partial_{\delta}\tilde{h}^{\mu\lambda}\partial_{\lambda}\tilde{h}^{\nu\sigma}
-g^{\sigma\delta}\partial_{\delta}\tilde{h}^{\mu\lambda}\partial_{\lambda}\tilde{h}^{\nu\rho}
+g^{\sigma\delta}\partial_{\delta}\tilde{h}^{\nu\lambda}\partial_{\lambda}\tilde{h}^{\mu\rho}\notag\\
&-g^{\rho\delta}\partial_{\delta}\tilde{h}^{\nu\lambda}\partial_{\lambda}\tilde{h}^{\mu\sigma}
+g^{\mu\delta}\partial_{\delta}\tilde{h}^{\rho\lambda}\partial_{\lambda}\tilde{h}^{\nu\sigma}
-g^{\mu\delta}\partial_{\delta}\tilde{h}^{\sigma\lambda}\partial_{\lambda}\tilde{h}^{\nu\rho}
+g^{\nu\delta}\partial_{\delta}\tilde{h}^{\sigma\lambda}\partial_{\lambda}\tilde{h}^{\mu\rho}
-g^{\nu\delta}\partial_{\delta}\tilde{h}^{\rho\lambda}\partial_{\lambda}\tilde{h}^{\mu\sigma}
+g^{\lambda\tau}\partial_{\lambda}\tilde{h}^{\mu\sigma}\partial_{\tau}\tilde{h}^{\nu\rho}\notag\\
&-g^{\lambda\tau}\partial_{\lambda}\tilde{h}^{\mu\rho}\partial_{\tau}\tilde{h}^{\nu\sigma}\big)
+\frac{1}{2}g_{\alpha\beta}\big(g^{\rho\epsilon}g^{\sigma\pi}\partial_{\epsilon}\tilde{h}^{\nu\beta}\partial_{\pi}\tilde{h}^{\mu\alpha}
-g^{\sigma\epsilon}g^{\rho\pi}\partial_{\epsilon}\tilde{h}^{\nu\beta}\partial_{\pi}\tilde{h}^{\mu\alpha}
+g^{\mu\epsilon}g^{\nu\pi}\partial_{\epsilon}\tilde{h}^{\sigma\beta}\partial_{\pi}\tilde{h}^{\rho\alpha}
-g^{\mu\epsilon}g^{\nu\pi}\partial_{\epsilon}\tilde{h}^{\rho\beta}\partial_{\pi}\tilde{h}^{\sigma\alpha}\notag\\
&+g^{\rho\epsilon}g^{\nu\pi}\partial_{\epsilon}\tilde{h}^{\mu\alpha}\partial_{\pi}\tilde{h}^{\sigma\beta}
-g^{\sigma\epsilon}g^{\nu\pi}\partial_{\epsilon}\tilde{h}^{\mu\alpha}\partial_{\pi}\tilde{h}^{\rho\beta}
+g^{\sigma\epsilon}g^{\mu\pi}\partial_{\epsilon}\tilde{h}^{\nu\alpha}\partial_{\pi}\tilde{h}^{\rho\beta}
-g^{\rho\epsilon}g^{\mu\pi}\partial_{\epsilon}\tilde{h}^{\nu\alpha}\partial_{\pi}\tilde{h}^{\sigma\beta}
+2g^{\rho\epsilon}g^{\nu\pi}\partial_{\epsilon}\tilde{h}^{\sigma\beta}\partial_{\pi}\tilde{h}^{\mu\alpha}\notag\\
&-2g^{\sigma\epsilon}g^{\nu\pi}\partial_{\epsilon}\tilde{h}^{\rho\beta}\partial_{\pi}\tilde{h}^{\mu\alpha}
+2g^{\sigma\epsilon}g^{\mu\pi}\partial_{\epsilon}\tilde{h}^{\rho\beta}\partial_{\pi}\tilde{h}^{\nu\alpha}
-2g^{\rho\epsilon}g^{\mu\pi}\partial_{\epsilon}\tilde{h}^{\sigma\beta}\partial_{\pi}\tilde{h}^{\nu\alpha}\big)
+\frac{1}{4}\big(g^{\sigma\delta}g^{\rho\nu}\partial_{\delta}\tilde{h}^{\mu\lambda}
-g^{\rho\delta}g^{\sigma\nu}\partial_{\delta}\tilde{h}^{\mu\lambda}\notag\\
&+g^{\rho\delta}g^{\sigma\mu}\partial_{\delta}\tilde{h}^{\nu\lambda}
-g^{\sigma\delta}g^{\rho\mu}\partial_{\delta}\tilde{h}^{\nu\lambda}
+g^{\mu\delta}g^{\rho\nu}\partial_{\delta}\tilde{h}^{\sigma\lambda}
-g^{\mu\delta}g^{\sigma\nu}\partial_{\delta}\tilde{h}^{\rho\lambda}
+g^{\nu\delta}g^{\sigma\mu}\partial_{\delta}\tilde{h}^{\rho\lambda}
-g^{\nu\delta}g^{\rho\mu}\partial_{\delta}\tilde{h}^{\sigma\lambda}\big)g_{\alpha\beta}\partial_{\lambda}\tilde{h}^{\alpha\beta}\notag\\
&+\frac{1}{4}\big(g^{\sigma\nu}\partial_{\lambda}\tilde{h}^{\rho\mu}-g^{\rho\nu}\partial_{\lambda}\tilde{h}^{\sigma\mu}
+g^{\rho\mu}\partial_{\lambda}\tilde{h}^{\sigma\nu}-g^{\sigma\mu}\partial_{\lambda}\tilde{h}^{\rho\nu}\big)g^{\lambda\tau}g_{\alpha\beta}\partial_{\tau}\tilde{h}^{\alpha\beta}
+\frac{1}{8}g_{\alpha\beta}g_{\lambda\tau}\big(g^{\rho\epsilon}g^{\mu\pi}g^{\sigma\nu}-g^{\sigma\epsilon}g^{\mu\pi}g^{\rho\nu}\notag\\
&+g^{\sigma\epsilon}g^{\nu\pi}g^{\rho\mu}
-g^{\rho\epsilon}g^{\nu\pi}g^{\sigma\mu}\big)\big(4\partial_{\epsilon}\tilde{h}^{\beta\tau}\partial_{\pi}\tilde{h}^{\alpha\lambda}
+\partial_{\epsilon}\tilde{h}^{\lambda\tau}\partial_{\pi}\tilde{h}^{\alpha\beta}\big)
+\frac{1}{8}\big(g^{\mu\sigma}g^{\rho\nu}-g^{\mu\rho}g^{\sigma\nu}\big)g^{\lambda\tau}g_{\alpha\beta}g_{\epsilon\pi}\partial_{\lambda}\tilde{h}^{\alpha\beta}\partial_{\tau}\tilde{h}^{\epsilon\pi}\bigg)\notag\\
&+\frac{1}{4}\big(g^{\mu\sigma}g^{\nu\rho}-g^{\mu\rho}g^{\nu\sigma}\big)g^{\lambda\tau}\partial_{\lambda}\ln{f_{R}}\partial_{\tau}\ln{f_{R}}
+\frac{1}{4}\big(g^{\rho\epsilon}g^{\mu\pi}g^{\sigma\nu}-g^{\sigma\epsilon}g^{\mu\pi}g^{\rho\nu}
+g^{\sigma\epsilon}g^{\nu\pi}g^{\rho\mu}-g^{\rho\epsilon}g^{\nu\pi}g^{\sigma\mu}\big)
\big(2\partial_{\epsilon}\partial_{\pi}\ln{f_{R}}\notag\\
&+\partial_{\epsilon}\ln{f_{R}}\partial_{\pi}\ln{f_{R}}\big)
-\frac{1}{4f_{R}\sqrt{-g}}\bigg(\big(g^{\sigma\delta}g^{\rho\nu}\partial_{\delta}\tilde{h}^{\mu\lambda}
-g^{\rho\delta}g^{\sigma\nu}\partial_{\delta}\tilde{h}^{\mu\lambda}
+g^{\rho\delta}g^{\sigma\mu}\partial_{\delta}\tilde{h}^{\nu\lambda}
-g^{\sigma\delta}g^{\rho\mu}\partial_{\delta}\tilde{h}^{\nu\lambda}
+g^{\mu\delta}g^{\rho\nu}\partial_{\delta}\tilde{h}^{\sigma\lambda}\notag\\
&-g^{\mu\delta}g^{\sigma\nu}\partial_{\delta}\tilde{h}^{\rho\lambda}
+g^{\nu\delta}g^{\sigma\mu}\partial_{\delta}\tilde{h}^{\rho\lambda}
-g^{\nu\delta}g^{\rho\mu}\partial_{\delta}\tilde{h}^{\sigma\lambda}\big)\partial_{\lambda}\ln{f_{R}}
+\big(g^{\sigma\nu}\partial_{\lambda}\tilde{h}^{\rho\mu}
-g^{\rho\nu}\partial_{\lambda}\tilde{h}^{\sigma\mu}
+g^{\rho\mu}\partial_{\lambda}\tilde{h}^{\sigma\nu}
-g^{\sigma\mu}\partial_{\lambda}\tilde{h}^{\rho\nu}\big)\times\notag\\
&g^{\lambda\tau}\partial{\tau}\ln{f_{R}}
+\frac{1}{2}g_{\alpha\beta}\big(g^{\rho\epsilon}g^{\mu\pi}g^{\sigma\nu}-g^{\sigma\epsilon}g^{\mu\pi}g^{\rho\nu}
+g^{\sigma\epsilon}g^{\nu\pi}g^{\rho\mu}-g^{\rho\epsilon}g^{\nu\pi}g^{\sigma\mu}\big)
\big(\partial_{\pi}\tilde{h}^{\alpha\beta}\partial{\epsilon}\ln{f_{R}}+\partial_{\epsilon}\tilde{h}^{\alpha\beta}\partial{\pi}\ln{f_{R}}\big)\notag\\
&+\big(g^{\mu\sigma}g^{\rho\nu}-g^{\mu\rho}g^{\sigma\nu}\big)g^{\lambda\tau}g_{\epsilon\pi}\partial_{\tau}\tilde{h}^{\epsilon\pi}\partial_{\lambda}\ln{f_{R}}\bigg),\\
R^{\mu\nu}&=-\frac{1}{2f_{R}\sqrt{-g}}g^{\alpha\beta}\partial_{\alpha}\tilde{h}^{\mu\nu}\partial_{\beta}\ln{f_{R}}+
\frac{1}{2f_{R}\sqrt{-g}}g^{\alpha\beta}\partial_{\alpha}\partial_{\beta}\tilde{h}^{\mu\nu}
-\frac{1}{2}g^{\mu\nu}g^{\alpha\beta}\partial_{\alpha}\ln{f_{R}}\partial_{\beta}\ln{f_{R}}
+\frac{1}{2}g^{\mu\alpha}g^{\nu\beta}\partial_{\alpha}\ln{f_{R}}\partial_{\beta}\ln{f_{R}}\notag\\
&+\frac{1}{2}g^{\mu\nu}g^{\alpha\beta}\partial_{\alpha}\partial_{\beta}\ln{f_{R}}+
g^{\mu\alpha}g^{\nu\beta}\partial_{\alpha}\partial_{\beta}\ln{f_{R}}
-\frac{1}{gf_{R}^{2}}g_{\beta\tau}g^{\alpha(\mu}\partial_{\lambda}\tilde{h}^{\nu)\tau}\partial_{\alpha}\tilde{h}^{\beta\lambda}
+\frac{1}{4gf_{R}^{2}}g_{\lambda\tau}g_{\epsilon\pi}g^{\mu\alpha}g^{\nu\beta}\partial_{\alpha}\tilde{h}^{\lambda\pi}
\partial_{\beta}\tilde{h}^{\tau\epsilon}\notag\\
&-\frac{1}{2f_{R}\sqrt{-g}}g_{\rho\sigma}g^{\mu(\alpha}g^{\beta)\nu}\partial_{\alpha}\tilde{h}^{\rho\sigma}\partial_{\beta}\ln{f_{R}}
+\frac{1}{2gf_{R}^{2}}g_{\alpha\beta}g^{\lambda\tau}\partial_{\lambda}\tilde{h}^{\mu\alpha}\partial_{\tau}\tilde{h}^{\nu\beta}
+\frac{1}{2gf_{R}^{2}}\partial_{\alpha}\tilde{h}^{\mu\beta}\partial_{\beta}\tilde{h}^{\nu\alpha}\notag\\
&+\frac{1}{f_{R}\sqrt{-g}}g^{\alpha(\mu}\partial_{\alpha}\tilde{h}^{\nu)\beta}\partial_{\beta}\ln{f_{R}}
-\frac{1}{8gf_{R}^{2}}g_{\tau\epsilon}g_{\lambda\pi}g^{\mu\alpha}g^{\nu\beta}\partial_{\alpha}\tilde{h}^{\lambda\pi}
\partial_{\beta}\tilde{h}^{\tau\epsilon}
-\frac{1}{4gf_{R}^{2}}g_{\rho\tau}g_{\epsilon\sigma}g^{\mu\nu}g^{\alpha\beta}\partial_{\alpha}\tilde{h}^{\rho\sigma}
\partial_{\beta}\tilde{h}^{\tau\epsilon}\notag
\end{align}
\begin{align}
\label{equ3.18}&-\frac{1}{4f_{R}\sqrt{-g}}g^{\mu\nu}g^{\alpha\beta}g_{\rho\sigma}\partial_{\alpha}\partial_{\beta}\tilde{h}^{\rho\sigma}
+\frac{1}{4f_{R}\sqrt{-g}}g^{\mu\nu}g^{\alpha\beta}g_{\rho\sigma}\partial_{\alpha}\tilde{h}^{\rho\sigma}\partial_{\beta}\ln{f_{R}},\\
R&=-\frac{1}{4gf_{R}^{2}}g_{\rho\tau}g_{\epsilon\sigma}g^{\alpha\beta}\partial_{\alpha}\tilde{h}^{\rho\sigma}\partial_{\beta}\tilde{h}^{\tau\epsilon}
-\frac{3}{2}g^{\alpha\beta}\partial_{\alpha}\ln{f_{R}}\partial_{\beta}\ln{f_{R}}
-\frac{1}{2f_{R}\sqrt{-g}}g_{\rho\sigma}g^{\alpha\beta}\partial_{\alpha}\partial_{\beta}\tilde{h}^{\rho\sigma}
+3g^{\alpha\beta}\partial_{\alpha}\partial_{\beta}\ln{f_{R}}\notag\\
\label{equ3.19}&-\frac{1}{2gf_{R}^{2}}g_{\alpha\beta}\partial_{\lambda}\tilde{h}^{\alpha\tau}\partial_{\tau}\tilde{h}^{\beta\lambda}
-\frac{1}{8gf_{R}^{2}}g^{\alpha\beta}g_{\tau\epsilon}g_{\lambda\pi}\partial_{\alpha}\tilde{h}^{\lambda\pi}\partial_{\beta}\tilde{h}^{\tau\epsilon}.
\end{align}
Further, the GB curvature scalar is
\begin{align}\label{equ3.20}
\mathcal{G}=&\frac{1}{2gf_{R}^2}\big(2g_{\lambda\alpha}g_{\tau\beta}-g_{\lambda\tau}g_{\alpha\beta}\big)\big(g^{\rho\nu}g^{\epsilon\pi}-g^{\rho\epsilon}g^{\nu\pi}\big)
\partial_{\rho}\partial_{\nu}\tilde{h}^{\lambda\tau}\partial_{\epsilon}\partial_{\pi}\tilde{h}^{\alpha\beta}
+\frac{2}{gf_{R}^2}g_{\sigma\tau}g^{\rho\epsilon}
\partial_{\rho}\partial_{\nu}\tilde{h}^{\pi\tau}\partial_{\epsilon}\partial_{\pi}\tilde{h}^{\nu\sigma}
-\frac{1}{gf_{R}^2}\partial_{\rho}\partial_{\nu}\tilde{h}^{\epsilon\pi}\partial_{\epsilon}\partial_{\pi}\tilde{h}^{\rho\nu}\notag\\
&+\frac{1}{gf_{R}^2}g_{\lambda\tau}g^{\rho\nu}
\partial_{\rho}\partial_{\nu}\tilde{h}^{\epsilon\pi}\partial_{\epsilon}\partial_{\pi}\tilde{h}^{\lambda\tau}
-\frac{2}{f_{R}\sqrt{-g}}g^{\rho\nu}\partial_{\rho}\partial_{\nu}\tilde{h}^{\epsilon\pi}\partial_{\epsilon}\partial_{\pi}\ln{f_{R}}
+2\big(g^{\rho\nu}g^{\epsilon\pi}-g^{\rho\epsilon}g^{\nu\pi}\big)\partial_{\rho}\partial_{\nu}\ln{f_{R}}\partial_{\epsilon}\partial_{\pi}\ln{f_{R}}\notag\\
&+\cdots,
\end{align}
where the omitted terms will produce the third-order or higher-order terms under the post-Minkowskian method.

We begin to consider the field equations (\ref{equ2.21}) for $f(R,\mathcal{G})$ gravity. $H^{\mu\nu}$ in Eq.~(\ref{equ2.22}) can be split into three parts,
\begin{align}
\label{equ3.21}H^{\mu\nu}&:=H^{\mu\nu}_{1}+H^{\mu\nu}_{2}+H^{\mu\nu}_{3},
\end{align}
where
\begin{align}
\label{equ3.22}H^{\mu\nu}_{1}:&=-\frac{1}{2}g^{\mu\nu}f+R^{\mu\nu}f_{R}+\frac{g^{\mu\nu}}{2}f_{\mathcal{G}}\mathcal{G},\\
\label{equ3.23}H^{\mu\nu}_{2}:&=(g^{\mu\nu}\square-\nabla^{\mu}\nabla^{\nu})f_{R},\\
\label{equ3.24}H^{\mu\nu}_{3}:&=2g^{\mu\nu}R\square f_{\mathcal{G}}-2R\nabla^{\mu}\nabla^{\nu}f_{\mathcal{G}}+4R^{\nu}_{\phantom{\nu}\lambda}\nabla^{\lambda}\nabla^{\mu}f_{\mathcal{G}}+
4R^{\mu}_{\phantom{\mu}\lambda}\nabla^{\lambda}\nabla^{\nu}f_{\mathcal{G}}-4g^{\mu\nu}R^{\alpha\beta}
\nabla_{\alpha}\nabla_{\beta}f_{\mathcal{G}}-4R^{\mu\nu}\square f_{\mathcal{G}}\notag\\
&\phantom{=}+4R^{\mu\rho\nu\sigma}\nabla_{\rho}\nabla_{\sigma}f_{\mathcal{G}}.
\end{align}
From Lagrangian \eqref{equ1.1}, the expression of $f_{\mathcal{G}}$ is
\begin{equation}\label{equ3.25}
f_{\mathcal{G}}=a_{2}+a_{12}R+2a_{22}\mathcal{G}+a_{112}R^{2}+2a_{122}R\mathcal{G}+3a_{222}\mathcal{G}^2+\cdots,
\end{equation}
and then, by the expressions of $f_{R}$, \eqref{equ3.12}, Eq.~(\ref{equ3.22}) reads
\begin{align}
\label{equ3.26}
H^{\mu\nu}_{1}=&G^{\mu\nu}+2a_{11}R^{\mu\nu}R-\frac{a_{11}}{2}g^{\mu\nu}R^{2}+\cdots,
\end{align}
where $G^{\mu\nu}$ is the Einstein tensor, and the omitted terms will produce the third-order or higher-order terms under the post-Minkowskian method.
By Eqs.~(\ref{equ3.18}) and (\ref{equ3.19}), the expression of $G^{\mu\nu}$ in terms of $\tilde{h}^{\mu\nu}$ can be obtained~\cite{Wu:2017vvm},
\begin{align}
G^{\mu\nu}&=-\frac{1}{2f_{R}\sqrt{-g}}g^{\alpha\beta}\partial_{\alpha}\tilde{h}^{\mu\nu}\partial_{\beta}\ln{f_{R}}+
\frac{1}{2f_{R}\sqrt{-g}}g^{\alpha\beta}\partial_{\alpha}\partial_{\beta}\tilde{h}^{\mu\nu}
+\frac{1}{4}g^{\mu\nu}g^{\alpha\beta}\partial_{\alpha}\ln{f_{R}}\partial_{\beta}\ln{f_{R}}
+\frac{1}{2}g^{\mu\alpha}g^{\nu\beta}\partial_{\alpha}\ln{f_{R}}\partial_{\beta}\ln{f_{R}}\notag\\
&\phantom{=}-g^{\mu\nu}g^{\alpha\beta}\partial_{\alpha}\partial_{\beta}\ln{f_{R}}+
g^{\mu\alpha}g^{\nu\beta}\partial_{\alpha}\partial_{\beta}\ln{f_{R}}
-\frac{1}{gf_{R}^{2}}g_{\beta\tau}g^{\alpha(\mu}\partial_{\lambda}\tilde{h}^{\nu)\tau}\partial_{\alpha}\tilde{h}^{\beta\lambda}
-\frac{1}{2f_{R}\sqrt{-g}}g_{\rho\sigma}g^{\mu(\alpha}g^{\beta)\nu}\partial_{\alpha}\tilde{h}^{\rho\sigma}\partial_{\beta}\ln{f_{R}}\notag\\
&\phantom{=}+\frac{1}{2gf_{R}^{2}}g_{\alpha\beta}g^{\lambda\tau}\partial_{\lambda}\tilde{h}^{\mu\alpha}\partial_{\tau}\tilde{h}^{\nu\beta}
+\frac{1}{2gf_{R}^{2}}\partial_{\alpha}\tilde{h}^{\mu\beta}\partial_{\beta}\tilde{h}^{\nu\alpha}
+\frac{1}{f_{R}\sqrt{-g}}g^{\alpha(\mu}\partial_{\alpha}\tilde{h}^{\nu)\beta}\partial_{\beta}\ln{f_{R}}
+\frac{1}{4gf_{R}^{2}}g^{\mu\nu}g_{\alpha\beta}\partial_{\lambda}\tilde{h}^{\alpha\tau}
\partial_{\tau}\tilde{h}^{\beta\lambda} \notag \\
\label{equ3.27}&\phantom{=}+\frac{1}{4f_{R}\sqrt{-g}}g^{\mu\nu}g^{\alpha\beta}g_{\rho\sigma}
\partial_{\alpha}\tilde{h}^{\rho\sigma}\partial_{\beta}\ln{f_{R}}
+\frac{1}{16gf_{R}^{2}}(2g^{\mu\alpha}g^{\nu\beta}-g^{\mu\nu}g^{\alpha\beta})(2g_{\lambda\tau}g_{\epsilon\pi}
-g_{\epsilon\tau}g_{\lambda\pi})\partial_{\alpha}\tilde{h}^{\lambda\pi}\partial_{\beta}\tilde{h}^{\tau\epsilon}.
\end{align}
For scalar $f_{R}$,
\begin{align*}
\nabla^{\mu}\nabla^{\nu}f_{R}&=g^{\mu\alpha}g^{\nu\beta}\partial_{\alpha}\partial_{\beta}f_{R}-\Gamma^{\lambda\mu\nu}\partial_{\lambda}f_{R},\\
\square f_{R}&=g^{\alpha\beta}\partial_{\alpha}\partial_{\beta}f_{R}-\Gamma^{\lambda}\partial_{\lambda}f_{R},
\end{align*}
where the expression of $\Gamma^{\lambda\mu\nu}$ is in Eq.~\eqref{equ3.4}, and $\Gamma^{\lambda}:=g^{\mu\nu}\Gamma^{\lambda}_{\mu\nu}$.
By use of (\ref{equ3.1}), (\ref{equ3.2}), (\ref{equ3.5}), (\ref{equ3.14}), (\ref{equ3.15}),
and (\ref{equ3.16}), Eq.~(\ref{equ3.4}) can be expressed as
\begin{align}
\Gamma^{\lambda\mu\nu}=&-\frac{1}{f_{R}\sqrt{-g}}g^{\rho(\mu}\partial_{\rho}\tilde{h}^{\nu)\lambda}+
\frac{1}{2f_{R}\sqrt{-g}}g^{\lambda\rho}\partial_{\rho}\tilde{h}^{\mu\nu}
+\frac{1}{2f_{R}\sqrt{-g}}g_{\alpha\beta}g^{\rho(\mu}g^{\nu)\lambda}\partial_{\rho}\tilde{h}^{\alpha\beta}
-\frac{1}{4f_{R}\sqrt{-g}}g_{\alpha\beta}g^{\mu\nu}g^{\lambda\rho}\partial_{\rho}\tilde{h}^{\alpha\beta} \notag\\
\label{equ3.29}\phantom{=}&-g^{\rho(\mu}g^{\nu)\lambda}\partial_{\rho}\ln{f_{R}}
+\frac{1}{2}g^{\mu\nu}g^{\lambda\rho}\partial_{\rho}\ln{f_{R}}.
\end{align}
Therefore,
\begin{align}
H^{\mu\nu}_{2}=&f_{R}g^{\mu\nu}g^{\alpha\beta}\partial_{\alpha}\partial_{\beta}\ln{f_{R}}-
f_{R}g^{\mu\alpha}g^{\nu\beta}\partial_{\alpha}\partial_{\beta}\ln{f_{R}}-f_{R}g^{\mu\alpha}g^{\nu\beta}\partial_{\alpha}\ln{f_{R}}\partial_{\beta}\ln{f_{R}}
-\frac{1}{\sqrt{-g}}g^{\rho(\mu}\partial_{\rho}\tilde{h}^{\nu)\lambda}\partial_{\lambda}\ln{f_{R}}\notag\\
\phantom{=}&+
\frac{1}{2\sqrt{-g}}g^{\lambda\rho}\partial_{\rho}\tilde{h}^{\mu\nu}\partial_{\lambda}\ln{f_{R}}
+\frac{1}{2\sqrt{-g}}g_{\alpha\beta}g^{\rho(\mu}g^{\nu)\lambda}\partial_{\rho}\tilde{h}^{\alpha\beta}\partial_{\lambda}\ln{f_{R}}
-\frac{1}{4\sqrt{-g}}g_{\alpha\beta}g^{\mu\nu}g^{\lambda\rho}\partial_{\rho}\tilde{h}^{\alpha\beta}\partial_{\lambda}\ln{f_{R}}\notag\\
\label{equ3.30}\phantom{=}&-f_{R}g^{\rho(\mu}g^{\nu)\lambda}\partial_{\rho}\ln{f_{R}}\partial_{\lambda}\ln{f_{R}}
+\frac{1}{2}f_{R}g^{\mu\nu}g^{\lambda\rho}\partial_{\rho}\ln{f_{R}}\partial_{\lambda}\ln{f_{R}}.
\end{align}
For scalar $f_{\mathcal{G}}$, similarly to $f_{R}$, there are
\begin{align}
\label{equ3.31}\nabla^{\mu}\nabla^{\nu}f_{\mathcal{G}}&=g^{\mu\alpha}g^{\nu\beta}\partial_{\alpha}\partial_{\beta}f_{\mathcal{G}}-\Gamma^{\lambda\mu\nu}\partial_{\lambda}f_{\mathcal{G}},\\
\label{equ3.32}\nabla_{\mu}\nabla_{\nu}f_{\mathcal{G}}&=\partial_{\mu}\partial_{\nu}f_{\mathcal{G}}-\Gamma^{\lambda}_{\mu\nu}\partial_{\lambda}f_{\mathcal{G}},\\
\label{equ3.33}\square f_{\mathcal{G}}&=g^{\alpha\beta}\partial_{\alpha}\partial_{\beta}f_{\mathcal{G}}
-\Gamma^{\lambda}\partial_{\lambda}f_{\mathcal{G}}.
\end{align}
By Eqs.~\eqref{equ3.17}---\eqref{equ3.20}, \eqref{equ3.29} and the expressions of $f_{R}$ and $f_{\mathcal{G}}$, namely Eqs.~(\ref{equ3.12}) and (\ref{equ3.25}), Eq.~(\ref{equ3.24}) reads
\begin{align}
\label{equ3.34}H^{\mu\nu}_{3}=&
2Rg^{\mu\nu}g^{\alpha\beta}\partial_{\alpha}\partial_{\beta}f_{\mathcal{G}}
-2Rg^{\mu\alpha}g^{\nu\beta}\partial_{\alpha}\partial_{\beta}f_{\mathcal{G}}
+4R^{\nu\alpha}g^{\mu\beta}\partial_{\alpha}\partial_{\beta}f_{\mathcal{G}}+
4R^{\mu\alpha}g^{\nu\beta}\partial_{\alpha}\partial_{\beta}f_{\mathcal{G}}
-4g^{\mu\nu}R^{\alpha\beta}\partial_{\alpha}\partial_{\beta}f_{\mathcal{G}}\notag\\
&-4R^{\mu\nu}g^{\alpha\beta}\partial_{\alpha}\partial_{\beta}f_{\mathcal{G}}
+4R^{\mu\rho\nu\sigma}\partial_{\rho}\partial_{\sigma}f_{\mathcal{G}}+\cdots,
\end{align}
where similarly to Eq.~\eqref{equ3.26}, the omitted terms will also produce the third-order or higher-order terms under the post-Minkowskian method.

The combination of Eqs.~(\ref{equ3.21}), (\ref{equ3.25})---(\ref{equ3.27}), (\ref{equ3.30}), and (\ref{equ3.34}) brings about the expression of $H^{\mu\nu}$:
\begin{align}
\label{equ3.35}H^{\mu\nu}=-\frac{1}{2gf_{R}^{2}}(\square_{\eta} \tilde{h}^{\mu\nu}-\Lambda^{\mu\nu})
\end{align}
with $\square_{\eta}:=\eta^{\mu\nu}\partial_{\mu}\partial_{\nu}$ and
\begin{align}
\label{equ3.36}\Lambda^{\mu\nu}=&\Lambda^{\mu\nu}_{GR}(\tilde{h}^{\alpha\beta})
-\frac{1}{2}(1+2f_{R})\tilde{g}^{\mu\nu}\tilde{g}^{\alpha\beta}\partial_{\alpha}\ln{f_{R}}\partial_{\beta}\ln{f_{R}}
-(1-4f_{R})\tilde{g}^{\mu\alpha}\tilde{g}^{\nu\beta}\partial_{\alpha}\ln{f_{R}}\partial_{\beta}\ln{f_{R}}
-2(f_{R}-1)\tilde{g}^{\mu\nu}\tilde{g}^{\alpha\beta}\partial_{\alpha}\partial_{\beta}\ln{f_{R}}\notag\\
&+2(f_{R}-1)\tilde{g}^{\mu\alpha}\tilde{g}^{\nu\beta}\partial_{\alpha}\partial_{\beta}\ln{f_{R}}
-2(1-f_{R})\tilde{g}^{\alpha(\mu}\partial_{\alpha}\tilde{h}^{\nu)\beta}\partial_{\beta}\ln{f_{R}}
-\frac{1}{2}(1-f_{R})\tilde{g}_{\rho\sigma}\tilde{g}^{\mu\nu}\tilde{g}^{\alpha\beta}\partial_{\alpha}\tilde{h}^{\rho\sigma}\partial_{\beta}\ln{f_{R}}\notag\\
&-(f_{R}-1)\tilde{g}^{\alpha\beta}\partial_{\alpha}\tilde{h}^{\mu\nu}\partial_{\beta}\ln{f_{R}}-(f_{R}-1)\tilde{g}_{\rho\sigma}\tilde{g}^{\mu(\alpha}\tilde{g}^{\beta)\nu}\partial_{\alpha}\tilde{h}^{\rho\sigma}\partial_{\beta}\ln{f_{R}}
+a_{11}\sqrt{-g}f_{R}\tilde{g}^{\mu\nu}R^{2}+4a_{11}gf_{R}^{2}R^{\mu\nu}R\notag\\
&-4a_{12}\big(\tilde{g}^{\mu\nu}\tilde{g}^{\alpha\beta}-\tilde{g}^{\mu\alpha}\tilde{g}^{\nu\beta}\big)R\partial_{\alpha}\partial_{\beta}R
-16a_{12}\sqrt{-g}f_{R}\tilde{g}^{\beta(\mu}R^{\nu)\alpha}\partial_{\alpha}\partial_{\beta}R
+8a_{12}\sqrt{-g}f_{R}\tilde{g}^{\mu\nu}R^{\alpha\beta}\partial_{\alpha}\partial_{\beta}R\notag\\
&+8a_{12}\sqrt{-g}f_{R}\tilde{g}^{\alpha\beta}R^{\mu\nu}\partial_{\alpha}\partial_{\beta}R
+8a_{12}gf_{R}^2R^{\mu\rho\nu\sigma}\partial_{\rho}\partial_{\sigma}R+\cdots,
\end{align}
where
\begin{align}
\label{equ3.37}\Lambda^{\mu\nu}_{GR}(\tilde{h}^{\alpha\beta})=&-\tilde{h}^{\alpha\beta}\partial_{\alpha}\partial_{\beta}\tilde{h}^{\mu\nu}
+\partial_{\alpha}\tilde{h}^{\mu\beta}\partial_{\beta}\tilde{h}^{\nu\alpha}
+\frac{1}{2}\tilde{g}^{\mu\nu}\tilde{g}_{\alpha\beta}\partial_{\lambda}\tilde{h}^{\alpha\tau}\partial_{\tau}\tilde{h}^{\beta\lambda}
-\tilde{g}^{\mu\alpha}\tilde{g}_{\beta\tau}\partial_{\lambda}\tilde{h}^{\nu\tau}\partial_{\alpha}\tilde{h}^{\beta\lambda}
-\tilde{g}^{\nu\alpha}\tilde{g}_{\beta\tau}\partial_{\lambda}\tilde{h}^{\mu\tau}\partial_{\alpha}\tilde{h}^{\beta\lambda}\notag\\
&+\tilde{g}_{\alpha\beta}\tilde{g}^{\lambda\tau}\partial_{\lambda}\tilde{h}^{\mu\alpha}\partial_{\tau}\tilde{h}^{\nu\beta}
+\frac{1}{8}(2\tilde{g}^{\mu\alpha}\tilde{g}^{\nu\beta}-\tilde{g}^{\mu\nu}\tilde{g}^{\alpha\beta})(2\tilde{g}_{\lambda\tau}\tilde{g}_{\epsilon\pi}-\tilde{g}_{\tau\epsilon}
\tilde{g}_{\lambda\pi})\partial_{\alpha}\tilde{h}^{\lambda\pi}\partial_{\beta}\tilde{h}^{\tau\epsilon},
\end{align}
and
\begin{align}
\label{equ3.38}\tilde{g}_{\mu\nu}:=\frac{1}{\sqrt{-g}f_{R}}g_{\mu\nu}
\end{align}
satisfies
\begin{align}
\label{equ3.39}\tilde{g}_{\mu\lambda}\tilde{g}^{\lambda\nu}=\delta^{\nu}_{\mu}.
\end{align}
In Eqs.~\eqref{equ3.36} and~\eqref{equ3.37}, the quantity $\Lambda^{\mu\nu}_{GR}(h^{\alpha\beta})$ is the corresponding term in GR with the replacement of the variables $h^{\alpha\beta}$ by $\tilde h^{\alpha\beta}$. It is easy to check that $\Lambda^{\mu\nu}$ is made of, at least, quadratic terms of the effective gravitational
field amplitude $\tilde{h}^{\mu\nu}$ and its first and second derivatives.
Moreover, $\Lambda^{\mu\nu}$ can reduce to the expression of $\Lambda^{\mu\nu}_{GR}(h^{\alpha\beta})$
when $f(R,\mathcal{G})$ gravity reduces to GR by Eqs.~(\ref{equ3.12}) and (\ref{equ3.15}).
Eqs.~\eqref{equ3.35} and~\eqref{equ3.36} show that the coupling constant $a_{12}$ appears in the expression of $H^{\mu\nu}$, which, from the Lagrangian~\eqref{equ1.1}, implies that the GB curvature scalar $\mathcal{G}$ should have nontrivial effect in the theory. According to the Eq.~(\ref{equ3.35}), the field equations of $f(R,\mathcal{G})$ gravity are transformed into the form of an obvious wave equation
\begin{equation}\label{equ3.40}
\square_{\eta} \tilde{h}^{\mu\nu}=2\kappa\tau^{\mu\nu}
\end{equation}
under the de Donder condition, where
the source terms
\begin{equation}\label{equ3.41}
\tau^{\mu\nu}=|g|f_{R}^{2}T^{\mu\nu}+\frac{1}{2\kappa}\Lambda^{\mu\nu}
\end{equation}
is the stress-energy pseudotensor of the matter fields and the gravitational field. Obviously,
Eqs.~(\ref{equ3.14}) and (\ref{equ3.40}) imply the conservation of the stress-energy pseudotensor in $f(R,\mathcal{G})$ gravity, namely,
\begin{equation}
\label{equ3.42}\partial_{\mu}\tau^{\mu\nu}=0.
\end{equation}
\subsection{Linearized $f(R,\mathcal{G})$ gravity}
If $\tilde{h}^{\mu\nu}$ is a perturbation, namely,
\begin{equation}\label{equ3.43}
|\tilde{h}^{\mu\nu}|\ll1,
\end{equation}
the linearized gravitational field equations are
\begin{align}
\label{equ3.44}\square_{\eta}\tilde{h}&^{\mu\nu}=2\kappa T^{\mu\nu}
\end{align}
by (\ref{equ3.40}) and (\ref{equ3.41}). Eq.~(\ref{equ3.44}) is the basis of multipole expansion of linearized $f(R,\mathcal{G})$ gravity with irreducible Cartesian tensors. It is easy to check that the field equations
(\ref{equ3.44}) of linearized $f(R,\mathcal{G})$ gravity are the same as those in linearized $f(R)$ gravity~\cite{Wu:2017vvm}. This is a trivial result because the leading term of ${\cal G}$ in the Lagrangian \eqref{equ1.1} gives a topological invariant, and its next leading term only relates to the higher-order effects.

Above field equations \eqref{equ3.44} are the linear results, but the effective stress-energy tensor of GWs for the linearized $f(R,\mathcal{G})$ gravity is the second-order nonlinear quantity, so in order to obtain its expression, we need some formulas to derive the quadratic terms of the stress-energy pseudotensor of the gravitational field. Firstly, by Eqs.~(\ref{equ3.10}) and (\ref{equ3.39}), we have
\begin{equation}\label{equ3.45}
\tilde{g}_{\mu\nu}=\eta_{\mu\nu}-\tilde{h}_{\mu\nu}.
\end{equation}
Next, by Eq.~(\ref{equ3.12}),
\begin{align}
\label{equ3.46}f_{R}&=1+2a_{11}R^{(1)}+o(\tilde{h}^{\mu\nu}),
\end{align}
where $o(\tilde{h}^{\mu\nu})$ is the higher-order terms of $\tilde{h}^{\mu\nu}$, and then inserting $f_{R}$ into
Eq.~(\ref{equ3.15}) gives rise to
\begin{align}
\label{equ3.47}h^{\mu\nu}&=\tilde{h}^{\mu\nu}-2a_{11}R^{(1)}\eta^{\mu\nu},\\
\label{equ3.48}h&=\tilde{h}-8a_{11}R^{(1)},
\end{align}
where $h=\eta_{\mu\nu}h^{\mu\nu}$ and $\tilde{h}=\eta_{\mu\nu}\tilde{h}^{\mu\nu}$. Eqs.~(\ref{equ3.47}) and (\ref{equ3.48}) show that $h^{\mu\nu}$ and $h$ are also perturbations, and thus, with the help of Eqs.~(\ref{equ3.1}) and (\ref{equ3.2}),
\begin{align}
\label{equ3.49}g&=|g_{\mu\nu}|=|\overline{g}^{\mu\nu}|=-1-h+o(h^{\mu\nu})=-1-\tilde{h}+8a_{11}R^{(1)}+o(\tilde{h}^{\mu\nu}),
\end{align}
so by Eq.~\eqref{equ3.46}, there are
\begin{align}
\label{equ3.50}\sqrt{-g}f_{R}&=1+\frac{\tilde{h}}{2}-2a_{11}R^{(1)}+o(\tilde{h}^{\mu\nu}),\\
\label{equ3.51}-gf_{R}^{2}&=1+\tilde{h}-4a_{11}R^{(1)}+o(\tilde{h}^{\mu\nu}).
\end{align}
The substitution of Eqs.~(\ref{equ3.10}), \eqref{equ3.17}---\eqref{equ3.19}, (\ref{equ3.45}), (\ref{equ3.46}), (\ref{equ3.50}), and (\ref{equ3.51}) in Eq.~(\ref{equ3.36}) gives rise to the quadratic term of $\Lambda^{\mu\nu}$:
\begin{align}\label{equ3.52}
\Lambda^{\mu\nu(2)}=&\Lambda^{\mu\nu(2)}_{f(R)}
-4a_{12}\eta^{\mu\nu}R^{(1)}\square_{\eta} R^{(1)}
+4a_{12}R^{(1)}\partial^{\mu}\partial^{\nu}R^{(1)}
-8a_{12}R^{\alpha\nu(1)}\partial^{\mu}\partial_{\alpha}R^{(1)}
-8a_{12}R^{\alpha\mu(1)}\partial^{\nu}\partial_{\alpha}R^{(1)}\notag\\
&+8a_{12}\eta^{\mu\nu}R^{\alpha\beta(1)}\partial_{\alpha}\partial_{\beta}R^{(1)}
+8a_{12}R^{\mu\nu(1)}\square_{\eta} R^{(1)}
-8a_{12}R^{\mu\rho\nu\sigma(1)}\partial_{\rho}\partial_{\sigma}R^{(1)},
\end{align}
where $\partial^{\mu}=\eta^{\mu\lambda}\partial_\lambda$. In Eq.~\eqref{equ3.52},
\begin{align}\label{equ3.53}
\Lambda^{\mu\nu(2)}_{f(R)}=&\Lambda^{\mu\nu(2)}_{GR}(\tilde h^{\alpha\beta})+a_{11}\eta^{\mu\nu}{R^{(1)}}^2-4a_{11}R^{\mu\nu(1)}R^{(1)}
-6a_{11}^{2}\eta^{\mu\nu}\partial^{\alpha}R^{(1)}\partial_{\alpha}R^{(1)}
+12a_{11}^{2}\partial^{\mu}R^{(1)}\partial^{\nu}R^{(1)}
\notag\\
&-8a_{11}^{2}\eta^{\mu\nu}R^{(1)}\square_{\eta} R^{(1)}
+8a_{11}^{2}R^{(1)}\partial^{\mu}\partial^{\nu}R^{(1)}
\end{align}
is the corresponding term in $f(R)$ gravity, and
\begin{align}\label{equ3.54}
\Lambda^{\mu\nu(2)}_{GR}(\tilde h^{\alpha\beta})&=-\tilde{h}^{\alpha\beta}\partial_{\alpha}\partial_{\beta}\tilde{h}^{\mu\nu}
+\frac{1}{2}\partial^{\mu}\tilde{h}_{\alpha\beta}\partial^{\nu}\tilde{h}^{\alpha\beta}
-\frac{1}{4}\partial^{\mu}\tilde{h}\partial^{\nu}\tilde{h}
+\partial_{\beta}\tilde{h}^{\mu\alpha}(\partial^{\beta}\tilde{h}^{\nu}_{\alpha}+\partial_{\alpha}\tilde{h}^{\nu\beta})
-\partial^{\mu}\tilde{h}_{\alpha\beta}\partial^{\alpha}\tilde{h}^{\nu\beta}\notag\\
&-\partial^{\nu}\tilde{h}_{\alpha\beta}\partial^{\alpha}\tilde{h}^{\mu\beta}+\eta^{\mu\nu}\Big(-\frac{1}{4}\partial_{\lambda}\tilde{h}_{\alpha\beta}\partial^{\lambda}\tilde{h}^{\alpha\beta}
+\frac{1}{8}\partial_{\lambda}\tilde{h}\partial^{\lambda}\tilde{h}
+\frac{1}{2}\partial_{\alpha}\tilde{h}_{\beta\lambda}\partial^{\beta}\tilde{h}^{\alpha\lambda}\Big)
\end{align}
is the quadratic term of  $\Lambda^{\mu\nu}_{GR}(\tilde h^{\alpha\beta})$.

We need to simplify $\Lambda^{\mu\nu(2)}$. By use of Eqs.~(\ref{equ3.10}), (\ref{equ3.11}), (\ref{equ3.14}), (\ref{equ3.38}), (\ref{equ3.45}), (\ref{equ3.46}), (\ref{equ3.50}), and (\ref{equ3.51}), Eqs.~(\ref{equ3.17}), (\ref{equ3.18}), and (\ref{equ3.19}) reduce to
\begin{align}
\label{equ3.55}R^{\mu\nu\rho\sigma(1)}&=\frac{1}{2}\big(\partial^{\sigma}\partial^{\nu}\tilde{h}^{\mu\rho}
-\partial^{\rho}\partial^{\nu}\tilde{h}^{\mu\sigma}+\partial^{\rho}\partial^{\mu}\tilde{h}^{\nu\sigma}
-\partial^{\sigma}\partial^{\mu}\tilde{h}^{\nu\rho}\big)
-\frac{1}{4}\big(\eta^{\mu\rho}\partial^{\sigma}\partial^{\nu}\tilde{h}
-\eta^{\mu\sigma}\partial^{\rho}\partial^{\nu}\tilde{h}
+\eta^{\nu\sigma}\partial^{\rho}\partial^{\mu}\tilde{h}
-\eta^{\nu\rho}\partial^{\sigma}\partial^{\mu}\tilde{h}\big)\notag\\
&\phantom{=}+a_{11}\big(\eta^{\mu\rho}\partial^{\sigma}\partial^{\nu}R^{(1)}
-\eta^{\mu\sigma}\partial^{\rho}\partial^{\nu}R^{(1)}
+\eta^{\nu\sigma}\partial^{\rho}\partial^{\mu}R^{(1)}
-\eta^{\nu\rho}\partial^{\sigma}\partial^{\mu}R^{(1)}\big),\\
\label{equ3.56}R^{\mu\nu(1)}&=\frac{1}{2}\square_{\eta}\tilde{h}^{\mu\nu}
-\frac{1}{4}\eta^{\mu\nu}\square_{\eta}\tilde{h}+a_{11}\eta^{\mu\nu}\square_{\eta} R^{(1)}+2a_{11}\partial^{\mu}\partial^{\nu}R^{(1)},\\
\label{equ3.57}R^{(1)}&=-\frac{1}{2}\square_{\eta}\tilde{h}+6a_{11}\square_{\eta} R^{(1)}
\end{align}
under the de Donder condition, and they satisfy
\begin{align*}
\partial_{\rho}\partial_{\sigma}R^{\mu\rho\nu\sigma(1)}&=\square_{\eta}R^{\mu\nu(1)}
-\frac{1}{2}\partial^{\mu}\partial^{\nu}R^{(1)},\\
\partial_{\mu}R^{\mu\nu(1)}&=\frac{1}{2}\partial^{\nu}R^{(1)},
\end{align*}
which can also be derived by the Bianchi identity, in fact. Furthermore, Eqs.~(\ref{equ2.21}) and (\ref{equ3.35}) lead to
\begin{align}
\label{equ3.58}H^{\mu\nu(1)}&=\frac{1}{2}\square_{\eta}\tilde{h}^{\mu\nu}=\kappa T^{\mu\nu(1)},\\
\label{equ3.59}H^{(1)}&=\frac{1}{2}\square_{\eta}\tilde{h}=\kappa T^{(1)},
\end{align}
so outside the sources, there are
\begin{align}
\label{equ3.60}R^{(1)}&=6a_{11}\square_{\eta} R^{(1)},\\
\label{equ3.61}R^{\mu\nu(1)}&=\frac{1}{6}\eta^{\mu\nu}R^{(1)}+2a_{11}\partial^{\mu}\partial^{\nu}R^{(1)}
\end{align}
by Eqs.~(\ref{equ3.57}) and (\ref{equ3.56}), respectively. Moreover, Eqs.~(\ref{equ3.57}) and (\ref{equ3.59}) imply
\begin{align}
\label{equ3.62}\square_{\eta} R^{(1)}-m^{2}R^{(1)}=m^{2}\kappa T^{(1)},
\end{align}
where
\begin{align}
\label{equ3.63}m^{2}:=\frac{1}{6a_{11}},
\end{align}
which shows that $R^{(1)}$ satisfies a massive KG equation with an external source, and it is the same as that in
linearized $f(R)$ gravity.

Now, we use the above formulas to simplify $\Lambda^{\mu\nu(2)}$. Firstly, applying Eqs.~\eqref{equ3.55}---\eqref{equ3.57} to Eqs.~\eqref{equ3.52} and~\eqref{equ3.53}  gives rise to
\begin{align}
\Lambda^{\mu\nu(2)}=&\Lambda^{\mu\nu(2)}_{f(R)}
+8a_{11}a_{12}\eta^{\mu\nu}\partial^{\alpha}\partial^{\beta}R^{(1)}\partial_{\alpha}\partial_{\beta}R^{(1)}
-\frac{2a_{12}}{9a_{11}}\eta^{\mu\nu}{\left(R^{(1)}\right)}^2
-16a_{11}a_{12}\partial^{\mu}\partial_{\alpha}R^{(1)}\partial^{\nu}\partial^{\alpha}R^{(1)}\notag\\
&+\frac{8a_{12}}{3}R^{(1)}\partial^{\mu}\partial^{\nu}R^{(1)}
-4a_{12}\partial_{\rho}\partial_{\sigma}R^{(1)}\Big(\big(
\partial^{\mu}\partial^{\nu}\tilde{h}^{\rho\sigma}
+\partial^{\rho}\partial^{\sigma}\tilde{h}^{\mu\nu}
-\partial^{\rho}\partial^{\nu}\tilde{h}^{\mu\sigma}
-\partial^{\mu}\partial^{\sigma}\tilde{h}^{\rho\nu}\big)\notag\\
\label{equ3.64}&-\frac{1}{2}\big(\eta^{\rho\sigma}\partial^{\mu}\partial^{\nu}\tilde{h}
+\eta^{\mu\nu}\partial^{\rho}\partial^{\sigma}\tilde{h}
-\eta^{\mu\sigma}\partial^{\rho}\partial^{\nu}\tilde{h}
-\eta^{\rho\nu}\partial^{\mu}\partial^{\sigma}\tilde{h}\big)\Big),
\end{align}
where
\begin{align}
\label{equ3.65}\Lambda^{\mu\nu(2)}_{f(R)}=&\Lambda^{\mu\nu(2)}_{GR}(\tilde h^{\alpha\beta})+12a_{11}^{2}\partial^{\mu}R^{(1)}\partial^{\nu}R^{(1)}
-\eta^{\mu\nu}\left(a_{11}{\left(R^{(1)}\right)}^2+6a_{11}^{2}\partial^{\alpha}R^{(1)}\partial_{\alpha}R^{(1)}\right).
\end{align}
Next, we will take the average $\big<\cdots\big>$ over a small spatial volume (several wavelengths)
surrounding each point, and the relevant rules for the average are~\cite{Berry:2011pb}
\begin{align}
\label{equ3.66}\big<\partial_{\mu}X\big>&=0,\\
\label{equ3.67}\big<A(\partial_{\mu}B)\big>&=-\big<(\partial_{\mu}A)B\big>,
\end{align}
where $X,A,B$ are three arbitrary quantities.  By use of Eqs.~(\ref{equ3.14}), (\ref{equ3.58})---(\ref{equ3.60}), (\ref{equ3.66}), and (\ref{equ3.67}), the average of Eq.~(\ref{equ3.64}) outside the sources reduces to
\begin{align}
\label{equ3.68}\big<\Lambda^{\mu\nu(2)}\big>=\big<\Lambda^{\mu\nu(2)}_{f(R)}\big>=\big<\Lambda^{\mu\nu(2)}_{GR}(\tilde h^{\alpha\beta})\big>+12a_{11}^{2}\big<\partial^{\mu}R^{(1)}\partial^{\nu}R^{(1)}\big>.
\end{align}
That is, after the average over a small spatial volume, the terms associated with the coupling constant $a_{12}$ in Eq.~\eqref{equ3.64} are cancelled each other, and the dependence on $a_{12}$ vanishes. It gives the effective stress-energy tensor of GWs in linearized $f(R,\mathcal{G})$ gravity:
\begin{align}
\label{equ3.69}t^{\mu\nu}:=\frac{1}{2\kappa}\big<\Lambda^{\mu\nu(2)}\big>=\frac{1}{2\kappa}\big<\Lambda^{\mu\nu(2)}_{f(R)}\big>=t^{\mu\nu}_{f(R)}
=t^{\mu\nu}_{GR}(\tilde h^{\alpha\beta})+\frac{6a_{11}^{2}}{\kappa}\big<\partial^{\mu}R^{(1)}\partial^{\nu}R^{(1)}\big>,
\end{align}
where
\begin{equation}
\label{equ3.70}t^{\mu\nu}_{GR}(\tilde h^{\alpha\beta})=\frac{1}{4\kappa}\Big<\partial^{\mu}\tilde{h}_{\alpha\beta}\partial^{\nu}\tilde{h}^{\alpha\beta}
-\frac{1}{2}\partial^{\mu}\tilde{h}\partial^{\nu}\tilde{h}\Big>
\end{equation}
is the effective stress-energy tensor of GWs under the de Donder condition in linearized GR with the replacement of the variables $h^{\alpha\beta}$ by $\tilde h^{\alpha\beta}$~\cite{Carroll2004}.

It should be noted that the above result only depends on the coupling constant $a_{11}$ in the Lagrangian \eqref{equ1.1}, and has nothing to do with the GB curvature scalar $\mathcal{G}$. In other words, the effective stress-energy tensor of GWs in linearized $f(R,\mathcal{G})$ gravity is the same as that in linearized $f(R)$ gravity~\cite{Berry:2011pb,Wu:2017vvm}. As mentioned before, for $f(R,\mathcal{G})$ gravity, the GB curvature scalar $\mathcal{G}$ plays an important role in the nonlinear effects in general. However, although the effective stress-energy tensor of GWs in $f(R,\mathcal{G})$ gravity is the typical second-order nonlinear quantity,
the above result shows that $\mathcal{G}$ does not contribute to it.
\section{Energy, momentum, and angular momentum carried by GWs in linearized $f(R,\mathcal{G})$ gravity~\label{Sec:EneggyMomentumAgular}}
\subsection{Multipole expansion of linearized $f(R,\mathcal{G})$ gravity~\label{Sec:MultipoleExpansion}}
As the further second-order nonlinear quantities, the energy, momentum, and angular momentum carried by GWs in linearized $f(R,\mathcal{G})$ gravity need to be discussed. For GR~\cite{Thorne:1980ru} and $f(R)$ gravity~\cite{Wu:2018hjx}, the fluxes of the energy and the momentum can be evaluated by using the above effective stress-energy tensor of GWs, but the flux of the angular momentum cannot~\cite{Thorne:1980ru}. This problem also exists in the $f(R,\mathcal{G})$ gravity. Therefore, a way, not depending on the effective stress-energy tensor of GWs, to deal with the angular momentum is needed. In this section, by following Refs.~\cite{Thorne:1980ru,Peters:1964zz,Wu:2018hjx}, we will calculate the energy, momentum, and angular momentum carried by GWs in linearized $f(R,\mathcal{G})$ gravity in a unified way. This unified way requires that the multipole expansion of the linearized $f(R,\mathcal{G})$ gravity is discussed firstly.

Eq.~(\ref{equ3.47}) shows that we should deal with the multipole expansions of tensor part $\tilde{h}^{\mu\nu}$ and of the scalar part associated with $R^{(1)}$, respectively. As mentioned in the preceding section, the linearized field equations (\ref{equ3.44}) of linearized $f(R,\mathcal{G})$ gravity are the same as those in linearized $f(R)$ gravity~\cite{Wu:2017vvm}. So, under the same gauge condition, namely the de Donder condition~\eqref{equ3.14}, the multipole expansions of $\tilde{h}^{\mu\nu}$ in these two gravitational models are also the same, and their expression is~\cite{Wu:2018hjx}
\begin{equation}\label{equ4.1}
\left\{\begin{array}{l}
\displaystyle\tilde{h}^{00}(t,\boldsymbol{x})= -\frac{4G}{c^{2}}\sum_{l=0}^{\infty}\frac{(-1)^{l}}{l!}\partial_{I_{l}}\left( \frac{\hat{M}_{I_{l}}(u)}{r}\right),\smallskip\\
\displaystyle\tilde{h}^{0i}(t,\boldsymbol{x})= \frac{4G}{c^{3}}\sum_{l=1}^{\infty}\frac{(-1)^{l}}{l!}\partial_{I_{l-1}}\left(\frac{\partial_{t}\hat{M}_{iI_{l-1}}(u)}{r}\right) +\frac{4G}{c^{3}}\sum_{l=1}^{\infty}\frac{(-1)^{l}l}{(l+1)!}\epsilon_{iab}\partial_{aI_{l-1}}\left( \frac{\hat{S}_{bI_{l-1}}(u)}{r}\right),\smallskip\\
\displaystyle\tilde{h}^{ij}(t,\boldsymbol{x})=-\frac{4G}{c^{4}}\sum_{l=2}^{\infty}\frac{(-1)^{l}}{l!}\partial_{I_{l-2}}\left(\frac{\partial_{t}^{2}\hat{M}_{ijI_{l-2}}(u)}{r}\right) -\frac{8G}{c^{4}}\sum_{l=2}^{\infty}\frac{(-1)^{l}l}{(l+1)!}\partial_{aI_{l-2}}\left(\frac{\epsilon_{ab(i}\partial_{|t|}\hat{S}_{j)bI_{l-2}}(u)}{r}\right),
\end{array}\right.
\end{equation}
where
\begin{equation}\label{equ4.2}
\left\{\begin{array}{l}
\displaystyle\hat{M}_{I_{l}}(u)=\frac{1}{c^{2}}\int d^{3}x'\left( \hat{X'}_{I_{l}}\left(\overline{T}^{00}_{l}(u,\boldsymbol{x}')+\overline{T}^{aa}_{l}(u,\boldsymbol{x}')\right)
-\frac{4(2l+1)}{c(l+1)(2l+3)}\hat{X'}_{aI_{l}}\partial_{t}\overline{T}^{0a}_{l+1}(u,\boldsymbol{x}') \right . \\
\displaystyle\qquad\qquad\qquad\qquad\qquad\qquad\qquad\left .
+\frac{2(2l+1)}{c^2(l+1)(l+2)(2l+5)}\hat{X'}_{abI_{l}}\partial_{t}^{2}\overline{T}^{ab}_{l+2}(u,\boldsymbol{x}')\right ),\smallskip\\
\displaystyle\hat{S}_{I_{l}}(u)=\frac{1}{c}\int d^{3}x'\left(\epsilon_{ab<i_{1}}\hat{X'}_{|a|i_{2}\cdots i_{l}>}\overline{T}^{0b}_{l}(u,\boldsymbol{x}') \right . \\
\displaystyle\qquad\qquad\qquad\qquad\qquad\qquad\qquad\left . -\frac{2l+1}{c(l+2)(2l+3)}\epsilon_{ab<i_{1}}\hat{X'}_{|ac|i_{2}\cdots i_{l}>}\partial_{t}\overline{T}^{cb}_{l+1}(u,\boldsymbol{x}')\right),\ l\geq1
\end{array}\right.
\end{equation}
are referred to as the mass-type and current-type source multipole moments, respectively~\cite{Blanchet:2013haa};
the symbol $(i|t|j)$ represents that $t$ is not the symmetric indices, the symbol $<i_{1}|a|i_{2}\cdots i_{l}>$ and $<i_{1}|ac|i_{2}\cdots i_{l}>$
represent that $a$ and $ac$ are not STF indices, respectively; and~\cite{Damour:1990gj}
\begin{align}
\label{equ4.3}\overline{T}^{\mu\nu}_{l}(u,\boldsymbol{x}'):=\frac{(2l+1)!!}{2^{l+1}l!}\int_{-1}^{1}(1-z^2)^{l}T^{\mu\nu}\Big(u+\frac{zr'}{c},
\boldsymbol{x}'\Big)dz.
\end{align}

The Eq.~(\ref{equ4.1}) can be simplified by adopting the further gauge. Under an infinitesimal coordinate transformation, $x'^{\mu}\rightarrow x^{\mu}+\varepsilon^{\mu}$,
\begin{equation}
\label{equ4.4}h'^{\mu\nu}=h^{\mu\nu}+\partial^{\mu}\varepsilon^{\nu}+\partial^{\nu}\varepsilon^{\mu}-\eta^{\mu\nu}\partial_{\alpha}\varepsilon^{\alpha}. \end{equation}
Moreover, $R^{(1)}$ is an invariant under the coordinate transformation, so from Eq.~(\ref{equ3.47}),
\begin{equation}
\label{equ4.5}\tilde{h}'^{\mu\nu}=\tilde{h}^{\mu\nu}+\partial^{\mu}\varepsilon^{\nu}+\partial^{\nu}\varepsilon^{\mu}-\eta^{\mu\nu}
\partial_{\alpha}\varepsilon^{\alpha},
\end{equation}
which shows that the transverse and traceless (TT) gauge can be applied to the effective gravitational field amplitude $\tilde{h}^{\mu\nu}$ like that in GR~\cite{Naf:2011za}. Therefore, Eq.~(\ref{equ4.1}) can be written as
\begin{equation}\label{equ4.6}
\left\{\begin{array}{l}
\displaystyle\tilde{h}^{00}_{TT}(t,\boldsymbol{x})= 0,\bigskip\\
\displaystyle\tilde{h}^{0i}_{TT}(t,\boldsymbol{x})= 0,\medskip\\
\displaystyle\tilde{h}^{ij}_{TT}(t,\boldsymbol{x})=-\frac{4G}{c^{4}}\left(\sum_{l=2}^{\infty}\frac{(-1)^{l}}{l!}\partial_{I_{l-2}}\bigg(\frac{\partial_{t}^{2}\hat{M}_{ijI_{l-2}}(u)}{r}\bigg)\right)^{TT} -\frac{8G}{c^{4}}\left(\sum_{l=2}^{\infty}\frac{(-1)^{l}l}{(l+1)!}\partial_{aI_{l-2}}\bigg(\frac{\epsilon_{ab(i}\partial_{|t|}\hat{S}_{j)bI_{l-2}}(u)}{r}\bigg)\right)^{TT}
\end{array}\right.
\end{equation}
under the TT gauge, where the symbol $TT$ represents the TT projection operator which is defined as
\begin{equation}
\label{equ4.7}(X_{ij})^{TT}:=P_{ik}P_{jl}X_{kl}-\frac{1}{2}P_{ij}P_{kl}X_{kl},
\end{equation}
where $X_{ij}$ is an arbitrary symmetric spatial tensor, and
\begin{equation}
\label{equ4.8}P_{ij}=\delta_{ij}-n_{i}n_{j}.
\end{equation}
According to Refs.~\cite{Carroll2004,Maggiore2008}, $t^{\mu\nu}_{GR}(\tilde h^{\alpha\beta})$ in Eq.~\eqref{equ3.70}
can also be simplified by imposing the TT gauge outside the source, namely,
\begin{equation}
\label{equ4.9}t^{\mu\nu}_{GR}\big(\tilde h^{\alpha\beta}\big)=\frac{1}{4\kappa}\Big<\partial^{\mu}\tilde{h}_{pq}^{TT}\partial^{\nu}\tilde{h}_{pq}^{TT}\Big>.
\end{equation}

$R^{(1)}$ in linearized $f(R,\mathcal{G})$ gravity satisfies the massive KG equation~\eqref{equ3.62} with an external source, which is the same as that in linearized $f(R)$ gravity~\cite{Wu:2017vvm}, so in these two gravitational models, $R^{(1)}$ has the same multipole expansion. Thus, the multipole expansion of $R^{(1)}$ under the condition $r'/r\lll 1$ is~\cite{Wu:2018hjx},
\begin{align}
\label{equ4.10}R^{(1)}(t,\boldsymbol{x})
&=-\frac{m^{2}\kappa}{4\pi}\sum_{l=0}^{\infty}\frac{(-1)^l}{l!}\int d^{3}x'\frac{c}{r'}\hat{X}'_{I_{l}}(\theta',\varphi')\hat{T}_{I_{l}}(t,\boldsymbol{x},\boldsymbol{x}'),
\end{align}
where $r=|\boldsymbol{x}|$, $r'=|\boldsymbol{x}'|$,
\begin{align}
\label{equ4.11}
X'_{I_{l}}(\theta',\varphi')=&X'_{i_{1}i_{2}\cdots i_{l}}(\theta',\varphi'):= x'_{i_{1}}x'_{i_{2}}\cdots x'_{i_{l}}={r'}^l N_{I_l}(\theta',\varphi'),\\
\label{equ4.12}\hat{T}_{I_{l}}(t,\boldsymbol{x},\boldsymbol{x}')=&\frac{(2l+1)!!}{2^{l+1}l!}\left(\int_{u-\frac{r'}{c}}^{u+\frac{r'}{c}}dt'\left[\hat{\partial}_{I_{l}}\left(\frac{1}{r}\Big(1-\frac{c^2}{r'^2}(t'-u)^2\Big)^l\right)\right]
T^{(1)}(t',\boldsymbol{x}')\right.\notag\\
&\left.-\int_{-\infty}^{u-\frac{r'}{c}}dt'\left[\hat{A}(t,r;t',r')\hat{\partial}_{I_{l}}\left(\frac{1}{r}\Big(1-\frac{c^2}{r'^2}(t'-u)^2\Big)^l\right)\right]
T^{(1)}(t',\boldsymbol{x}')\right)
\end{align}
with
\begin{align}
\label{equ4.13}\hat{A}(t,r;t',r'):&=\exp{\Big(\frac{m^2rr'}{2}(\text{int}\ t')_{-}\Big)}-\exp{\Big(\frac{m^2rr'}{2}(\text{int}\ t')_{+}\Big)},\\
\label{equ4.14}(\text{int}\ t')_{\pm}\tilde{f}(t'):&=-\frac{c}{r'}\int_{t'}^{u\pm\frac{r'}{c}}\tilde{f}(\tau)\Big(1+\frac{c}{r}(u-\tau)\Big)d\tau.
\end{align}
Upon obtaining Eqs.~(\ref{equ4.1}) and (\ref{equ4.10}), we can write down the multipole expansion of $h^{\mu\nu}$ under the de Donder condition by Eq.~(\ref{equ3.47}). Under the condition $r'/r\lll 1$, it is
\begin{equation}\label{equ4.15}
\left\{\begin{array}{l}
\displaystyle h^{00}(t,\boldsymbol{x})= -\frac{4G}{c^{2}}\sum_{l=0}^{\infty}\frac{(-1)^{l}}{l!}\partial_{I_{l}}\left(\frac{\hat{M}_{I_{l}}(u)}{r}\right)
-\frac{2G}{3c^4}\sum_{l=0}^{\infty}\frac{(-1)^l}{l!}\int d^{3}x'\frac{c}{r'}\hat{X}'_{I_{l}}(\theta',\varphi')\hat{T}_{I_{l}}(t,\boldsymbol{x},\boldsymbol{x}'),\\
\displaystyle h^{0i}(t,\boldsymbol{x})= \frac{4G}{c^{3}}\sum_{l=1}^{\infty}\frac{(-1)^{l}}{l!}\partial_{I_{l-1}}\left(\frac{\partial_{t}\hat{M}_{iI_{l-1}}(u)}{r}\right)+\frac{4G}{c^{3}}\sum_{l=1}^{\infty}\frac{(-1)^{l}l}{(l+1)!}\epsilon_{iab}\partial_{aI_{l-1}}\left(\frac{\hat{S}_{bI_{l-1}}(u)}{r}\right),\\
\displaystyle h^{ij}(t,\boldsymbol{x})=-\frac{4G}{c^{4}}\sum_{l=2}^{\infty}\frac{(-1)^{l}}{l!}\partial_{I_{l-2}}\left(\frac{\partial_{t}^{2}\hat{M}_{ijI_{l-2}}(u)}{r}\right) -\frac{8G}{c^{4}}\sum_{l=2}^{\infty}\frac{(-1)^{l}l}{(l+1)!}\partial_{aI_{l-2}}\left(\frac{\epsilon_{ab(i}\partial_{|t|}\hat{S}_{j)bI_{l-2}}(u)}{r}\right)\\
\displaystyle \phantom{h^{00}(t,\boldsymbol{x})=}+\frac{2G}{3c^4}\delta^{ij}\sum_{l=0}^{\infty}\frac{(-1)^l}{l!}\int d^{3}x'\frac{c}{r'}\hat{X}'_{I_{l}}(\theta',\varphi')\hat{T}_{I_{l}}(t,\boldsymbol{x},\boldsymbol{x}'),
\end{array}\right.
\end{equation}
which is the same as that in the linearized $f(R)$ gravity~\cite{Wu:2017vvm}, as expected.

As to the multipole expansion of $h^{\mu\nu}$ in the radiation field, we still need to deal with the tensor part $\tilde{h}^{\mu\nu}$ and the scalar part associated with $R^{(1)}$, separately.  By Eq.~\eqref{equ2.15}, we have
\begin{align}
\label{equ4.16}&\hat{\partial}_{I_{l}}\left(\frac{F(u)}{r}\right)=\hat{N}_{I_{l}}(\theta,\varphi)\frac{(-1)^{l}}{c^{l}}
\frac{\partial_{u}^{l}F(u)}{r}+O\left(\frac{1}{r^2}\right),
\end{align}
where $O(1/r^2)$ contains the terms whose orders are equal to or higher than $1/r^2$.
The $1/r$ expansion in the distance to the source requires formula~\eqref{equ4.16} to be applied to Eqs.~\eqref{equ4.6} and~\eqref{equ4.10}, and then the multipole expansions of $\tilde{h}^{\mu\nu}$ and $R^{(1)}$ are obtained, respectively,
\begin{equation}\label{equ4.17}
\left\{\begin{array}{l}
\displaystyle\tilde{h}^{00}_{TT}(t,\boldsymbol{x})= 0,\bigskip\\
\displaystyle\tilde{h}^{0i}_{TT}(t,\boldsymbol{x})= 0,\medskip\\
\displaystyle\tilde{h}^{ij}_{TT}(t,\boldsymbol{x})=-\frac{4G}{c^{2}r}\sum_{l=2}^{\infty}\left(\frac{1}{c^{l}l!}
\Big(\partial_{u}^{l}\hat{M}_{ijI_{l-2}}(u)N_{I_{l-2}}(\theta,\varphi)\Big)^{TT} -\frac{2l}{c^{l+1}(l+1)!}\Big(\epsilon^{}_{ab(i}\partial_{|u|}^{l}\hat{S}_{j)bI_{l-2}}(u)N_{aI_{l-2}}(\theta,\varphi)\Big)^{TT}\right)\\
\displaystyle\qquad\qquad\qquad+O\left(\frac{1}{r^2}\right)
\end{array}\right.
\end{equation}
and
\begin{align}
\label{equ4.18}R^{(1)}(t,\boldsymbol{x})&=-\frac{m^{2}\kappa}{4\pi r}\sum_{l=0}^{\infty}\frac{1}{c^{l}l!}\hat{K}_{I_{l}}(t,r)N_{I_{l}}(\theta,\varphi)+O\left(\frac{1}{r^2}\right),
\end{align}
where
\begin{align}
\label{equ4.19}\hat{K}_{I_{l}}(t,r)=&\frac{(2l+1)!!}{2^{l+1}l!}\int d^{3}x'\frac{c}{r'}\hat{X'}_{I_{l}}(\theta',\varphi')\left(\int_{u-\frac{r'}{c}}^{u+\frac{r'}{c}}dt'\bigg[\frac{d^l}{du^l}\Big(1-\frac{c^2}{r'^2}(t'-u)^2\Big)^l\bigg]
T^{(1)}(t',\boldsymbol{x}')\right.\notag\\
&-\left.\int_{-\infty}^{u-\frac{r'}{c}}dt'\left[\hat{A}(t,r;t',r')\frac{d^l}{du^l}\Big(1-\frac{c^2}{r'^2}(t'-u)^2\Big)^l\right]T^{(1)}(t',\boldsymbol{x}')\right)
\end{align}
is the effective $l$-pole radiative moment of $R^{(1)}$. Eqs.~\eqref{equ4.17} and~\eqref{equ4.18}, of course, are the same as those in the linearized $f(R)$ gravity, respectively~\cite{Wu:2018hjx}.

According to the above conclusions, Eq.~\eqref{equ3.47} implies that $h^{\mu\nu}$ in linearized $f(R,\mathcal{G})$ gravity and in linearized $f(R)$ gravity have the same multipole expansion under the TT gauge in the radiation field, and it is~\cite{Wu:2018hjx}
\begin{equation}\label{equ4.20}
\left\{\begin{array}{l}
\displaystyle h^{00}(t,\boldsymbol{x})=-\frac{2G}{3c^4 r}\sum_{l=0}^{\infty}\frac{1}{c^{l}l!}\hat{K}_{I_{l}}(t,r)N_{I_{l}}(\theta,\varphi)+O\left(\frac{1}{r^2}\right),\smallskip\\
\displaystyle h^{0i}(t,\boldsymbol{x})= 0,\medskip\\
\displaystyle h^{ij}(t,\boldsymbol{x})=-\frac{4G}{c^{2}r}\sum_{l=2}^{\infty}\left(\frac{1}{c^{l}l!}\Big(\partial_{u}^{l}\hat{M}_{ijI_{l-2}}(u)N_{I_{l-2}}(\theta,\varphi)\Big)^{TT} -\frac{2l}{c^{l+1}(l+1)!}\Big(\epsilon^{}_{ab(i}\partial_{|u|}^{l}\hat{S}_{j)bI_{l-2}}(u)N_{aI_{l-2}}(\theta,\varphi)\Big)^{TT}\right)\medskip\\
\displaystyle \phantom{h^{ij}(t,\boldsymbol{x})=}+\frac{2G}{3c^4 r}\delta^{ij}\sum_{l=0}^{\infty}\frac{1}{c^{l}l!}\hat{K}_{I_{l}}(t,r)N_{I_{l}}(\theta,\varphi)+O\left(\frac{1}{r^2}\right).
\end{array}\right.
\end{equation}
\subsection{Energy, momentum, and angular momentum carried by GWs in linearized $f(R,\mathcal{G})$ gravity}
For $f(R,\mathcal{G})$ gravity, the energy, momentum, and angular momentum carried by the gravitational field inside a volume $V$ enclosed by a large sphere $S$ are
\begin{align}
\label{equ4.21}E&=\int_{V}d^{3}x\ \tau^{00},\\
\label{equ4.22}P_{a}&=\frac{1}{c}\int_{V}d^{3}x\ \tau^{0a},\\
\label{equ4.23}J_{a}&=\frac{1}{c}\int_{V}d^{3}x\ \epsilon_{abc}x_{b}\tau^{0c},
\end{align}
respectively, and then, by Eq.~(\ref{equ3.42}) and the Gauss theorem, there are
\begin{align}
\label{equ4.24}\frac{dE}{dt}&=c\int_{V}d^{3}x\,\partial_{0}\tau^{00}=-c\int_{V}d^{3}x\,\partial_{i}\tau^{i0}=-c\int
d\Omega\,r^2n_{i}\tau^{i0},\\
\label{equ4.25}\frac{dP_{a}}{dt}&=\int_{V}d^{3}x\,\partial_{0}\tau^{0a}=-\int_{V}d^{3}x\,\partial_{i}\tau^{ia}=-\int d\Omega\,r^2n_{i}\tau^{ia},\\
\label{equ4.26}\frac{dJ_{a}}{dt}&=\int_{V}d^{3}x\,\partial_{0}(\epsilon_{abc}x_{b}\tau^{0c})
=\int_{V}d^{3}x\,\epsilon_{abc}x_{b}\partial_{0}\tau^{0c}
=-\int_{V}d^{3}x\,\epsilon_{abc}x_{b}\partial_{i}\tau^{ic}\notag\\
&=-\int_{V}d^{3}x\,\partial_{i}(\epsilon_{abc}x_{b}\tau^{ic})
=-\int d\Omega\,r^3\epsilon_{abc}n_{i}n_{b}\tau^{ic},
\end{align}
where $n_{i}$ is also the unit normal vector of the large sphere $S$, and $d\Omega$ is the element of solid angle whose integral domain is $4\pi$. Thus, for the linearized $f(R,\mathcal{G})$ gravity, the fluxes of energy, momentum, and angular momentum carried by the outward-propagating GW should be~\cite{Thorne:1980ru,Peters:1964zz}
\begin{align}
\label{equ4.27}\frac{dE}{d\Omega dt}&=+cr^2\Big<n_{i}(\tau^{i0}_{\rm{rad}})^{(2)}\Big>,\\
\label{equ4.28}\frac{dP_{a}}{d\Omega dt}&=+r^2\Big<n_{i}(\tau^{ia}_{\rm{rad}})^{(2)}\Big>,\\
\label{equ4.29}\frac{dJ_{a}}{d\Omega dt}&=+r^3\Big<\epsilon_{abc}n_{i}n_{b}(\tau^{ic}_{\rm{rad}})^{(2)}\Big>,
\end{align}
respectively, where $$(\tau^{\mu\nu}_{\rm{rad}})^{(2)}=\frac{1}{2\kappa}\left(\Lambda^{\mu\nu(2)}\right)$$ is the quadratic term of the stress-energy pseudotensor under the TT gauge in the radiation field.

Next, the leading terms of the fluxes of energy, momentum, and angular momentum carried by GWs in linearized $f(R,\mathcal{G})$ gravity are expressed by
\begin{align}
\label{equ4.30}\left(\frac{dE}{d\Omega dt}\right)^{[0]}&=cr^2\Big<n_{i}(\tau^{i0}_{\rm{rad}})^{(2)}\Big>^{[2]},\\
\label{equ4.31}\left(\frac{dP_{a}}{d\Omega dt}\right)^{[0]}&=r^2\Big<n_{i}(\tau^{ia}_{\rm{rad}})^{(2)}\Big>^{[2]},\\
\label{equ4.32}\left(\frac{dJ_{a}}{d\Omega dt}\right)^{[0]}&=r^3\Big<\epsilon_{abc}n_{i}n_{b}(\tau^{ic}_{\rm{rad}})^{(2)}\Big>^{[3]},
\end{align}
respectively, where the superscript $[m]\ (m=0,1,2,3\cdots)$ represents that the term of the $1/r^m$ order for the corresponding quantity is taken, and then, there are
\begin{align}
\label{equ4.33}\frac{dE}{d\Omega dt}&=\left(\frac{dE}{d\Omega dt}\right)^{[0]}+O\left(\frac{1}{r}\right),\\
\label{equ4.34}\frac{dP_{a}}{d\Omega dt}&=\left(\frac{dP_{a}}{d\Omega dt}\right)^{[0]}+O\left(\frac{1}{r}\right),\\
\label{equ4.35}\frac{dJ_{a}}{d\Omega dt}&=\left(\frac{dJ_{a}}{d\Omega dt}\right)^{[0]}+O\left(\frac{1}{r}\right).
\end{align}
In the following, the fluxes of the energy, momentum, and the angular momentum
are used to denote their leading terms~\eqref{equ4.30}---\eqref{equ4.32}, respectively. Moreover, we also need to prove
\begin{align}
\label{equ4.36}\Big<\epsilon_{abc}n_{i}n_{b}(\tau^{ic}_{\rm{rad}})^{(2)}\Big>^{[2]}=0.
\end{align}
Because if it does not hold, Eq.~(\ref{equ4.32}) shows that the flux of angular momentum carried by the outward-propagating GW will diverge, and this is impossible.

By Eqs.~(\ref{equ3.64}), (\ref{equ3.65}), and (\ref{equ3.54}), $(\tau^{i0}_{\rm{rad}})^{(2)}$ and $(\tau^{ij}_{\rm{rad}})^{(2)}$ under the TT gauge are, respectively,
\begin{align}
\label{equ4.37}(\tau^{i0}_{\rm{rad}})^{(2)}=&(\tau^{i0}_{\rm{rad}})^{(2)}_{f(R)}+\frac{1}{\kappa}\bigg(
-\frac{4a_{12}}{3}R^{(1)}\partial_{0}\partial_{i}R^{(1)}
+8a_{11}a_{12}\partial_{i}\partial_{\alpha}R^{(1)}\partial_{0}\partial^{\alpha}R^{(1)}
+2a_{12}\partial_{p}\partial_{q}R^{(1)}\partial_{0}\partial_{i}\tilde{h}_{pq}^{TT}\notag\\
&-2a_{12}\partial_{\rho}\partial_{q}R^{(1)}\partial^{\rho}\partial_{0}\tilde{h}_{iq}^{TT}\bigg),\\
\label{equ4.38}(\tau^{ij}_{\rm{rad}})^{(2)}=&(\tau^{ij}_{\rm{rad}})^{(2)}_{f(R)}+\frac{1}{\kappa}\bigg(
-\frac{a_{12}}{9a_{11}}\delta_{ij}{\left(R^{(1)}\right)}^2+4a_{11}a_{12}\delta_{ij}\partial^{\alpha}\partial^{\beta}R^{(1)}\partial_{\alpha}\partial_{\beta}R^{(1)}
+\frac{4a_{12}}{3}R^{(1)}\partial_{i}\partial_{j}R^{(1)}\notag\\
&-8a_{11}a_{12}\partial_{i}\partial_{\alpha}R^{(1)}\partial_{j}\partial^{\alpha}R^{(1)}
-2a_{12}\partial_{p}\partial_{q}R^{(1)}\partial_{i}\partial_{j}\tilde{h}_{pq}^{TT}
-2a_{12}\partial_{\rho}\partial_{\sigma}R^{(1)}\partial^{\rho}\partial^{\sigma}\tilde{h}_{ij}^{TT}\notag\\
&+2a_{12}\partial_{\rho}\partial_{q}R^{(1)}\partial^{\rho}\partial_{j}\tilde{h}_{iq}^{TT}
+2a_{12}\partial_{\sigma}\partial_{q}R^{(1)}\partial^{\sigma}\partial_{i}\tilde{h}_{jq}^{TT}\bigg),
\end{align}
where
\begin{align}
\label{equ4.39}(\tau^{i0}_{\rm{rad}})^{(2)}_{f(R)}=&\frac{1}{2\kappa}\bigg(-\frac{1}{2}\partial_{0}\tilde{h}_{pq}^{TT}\partial_{i}\tilde{h}_{pq}^{TT}
+\partial_{0}\tilde{h}_{pq}^{TT}\partial_{p}\tilde{h}_{iq}^{TT}-12a_{11}^{2}\partial_{0}R^{(1)}\partial_{i}R^{(1)}\bigg),\\
\label{equ4.40}(\tau^{ij}_{\rm{rad}})^{(2)}_{f(R)}=&\frac{1}{2\kappa}\bigg(-\tilde{h}_{pq}^{TT}\partial_{p}\partial_{q}\tilde{h}_{ij}^{TT}
+\frac{1}{2}\partial_{i}\tilde{h}_{pq}^{TT}\partial_{j}\tilde{h}_{pq}^{TT}
-\partial_{0}\tilde{h}_{ip}^{TT}\partial_{0}\tilde{h}_{jp}^{TT}+\partial_{q}\tilde{h}_{ip}^{TT}\partial_{q}\tilde{h}_{jp}^{TT}+\partial_{p}\tilde{h}_{iq}^{TT}\partial_{q}\tilde{h}_{jp}^{TT}
-\partial_{i}\tilde{h}_{pq}^{TT}\partial_{p}\tilde{h}_{jq}^{TT}\notag\\
&-\partial_{j}\tilde{h}_{pq}^{TT}\partial_{p}\tilde{h}_{iq}^{TT}+\delta_{ij}\Big(\frac{1}{4}\partial_{0}\tilde{h}_{pq}^{TT}\partial_{0}\tilde{h}_{pq}^{TT}-\frac{1}{4}\partial_{m}\tilde{h}_{pq}^{TT}\partial_{m}\tilde{h}_{pq}^{TT}
+\frac{1}{2}\partial_{p}\tilde{h}_{qm}^{TT}\partial_{q}\tilde{h}_{pm}^{TT}\Big)
+12a_{11}^{2}\partial_{i}R^{(1)}\partial_{j}R^{(1)}\notag\\
&-\delta_{ij}\Big(a_{11}{\left(R^{(1)}\right)}^2+6a_{11}^2\partial^{\alpha}R^{(1)}\partial_{\alpha}R^{(1)}\Big)\bigg),
\end{align}
and then, we can use them to calculate the average parts in Eqs. \eqref{equ4.30}---\eqref{equ4.32}, and \eqref{equ4.36}.
Firstly, the average parts in Eqs.~\eqref{equ4.30} and~\eqref{equ4.31} are, respectively,
\begin{align}
\label{equ4.41}\Big<n_{i}(\tau^{i0}_{\rm{rad}})^{(2)}\Big>^{[2]}&=\Big<n_{i}(\tau^{i0}_{\rm{rad}})^{(2)}_{f(R)}\Big>^{[2]}
+\frac{1}{\kappa}\bigg(-\frac{4a_{12}}{3}\Big<n_{i}R^{(1)}\partial_{0}\partial_{i}R^{(1)}\Big>^{[2]}
+8a_{11}a_{12}\Big<n_{i}\partial_{i}\partial_{\alpha}R^{(1)}\partial_{0}\partial^{\alpha}R^{(1)}\Big>^{[2]}\notag\\
&+2a_{12}\Big<n_{i}\partial_{p}\partial_{q}R^{(1)}\partial_{0}\partial_{i}\tilde{h}_{pq}^{TT}\Big>^{[2]}
-2a_{12}\Big<n_{i}\partial_{\rho}\partial_{q}R^{(1)}\partial^{\rho}\partial_{0}\tilde{h}_{iq}^{TT}\Big>^{[2]}\bigg),\\
\label{equ4.42}\Big<n_{i}(\tau^{ia}_{\rm{rad}})^{(2)}\Big>^{[2]}&=\Big<n_{i}(\tau^{ia}_{\rm{rad}})^{(2)}_{f(R)}\Big>^{[2]}
+\frac{1}{\kappa}\bigg(-\frac{a_{12}}{9a_{11}}\Big<n_{a}{\left(R^{(1)}\right)}^2\Big>^{[2]}
+4a_{11}a_{12}\Big<n_{a}\partial^{\alpha}\partial^{\beta}R^{(1)}\partial_{\alpha}\partial_{\beta}R^{(1)}\Big>^{[2]}\notag\\
&+\frac{4a_{12}}{3}\Big<n_{i}R^{(1)}\partial_{i}\partial_{a}R^{(1)}\Big>^{[2]}
-8a_{11}a_{12}\Big<n_{i}\partial_{i}\partial_{\alpha}R^{(1)}\partial_{a}\partial^{\alpha}R^{(1)}\Big>^{[2]}
-2a_{12}\Big<n_{i}\partial_{p}\partial_{q}R^{(1)}\partial_{i}\partial_{a}\tilde{h}_{pq}^{TT}\Big>^{[2]}\notag\\
&-2a_{12}\Big<n_{i}\partial_{\rho}\partial_{\sigma}R^{(1)}\partial^{\rho}\partial^{\sigma}\tilde{h}_{ia}^{TT}\Big>^{[2]}
+2a_{12}\Big<n_{i}\partial_{\rho}\partial_{q}R^{(1)}\partial^{\rho}\partial_{a}\tilde{h}_{iq}^{TT}\Big>^{[2]}
+2a_{12}\Big<n_{i}\partial_{\sigma}\partial_{q}R^{(1)}\partial^{\sigma}\partial_{i}\tilde{h}_{aq}^{TT}\Big>^{[2]}\bigg).
\end{align}
By using of Eqs.~\eqref{equ2.13}, \eqref{equ3.66}, \eqref{equ3.67}, \eqref{equ4.17}, and~\eqref{equ4.18},
\begin{align}
\label{equ4.43}\Big<n_{i}(\tau^{i0}_{\rm{rad}})^{(2)}\Big>^{[2]}&=\Big<n_{i}(\tau^{i0}_{\rm{rad}})^{(2)}_{f(R)}\Big>^{[2]}
+\frac{1}{\kappa}\bigg(-\frac{4a_{12}}{3}\Big<n_{i}R^{(1)}\partial_{0}\partial_{i}R^{(1)}\Big>^{[2]}
+8a_{11}a_{12}\Big<n_{i}R^{(1)}\partial_{0}\partial_{i}\square_{\eta}R^{(1)}\Big>^{[2]}\notag\\
&-2a_{12}\Big<n_{i}\partial_{p}R^{(1)}\partial_{0}\partial_{i}\partial_{q}\tilde{h}_{pq}^{TT}\Big>^{[2]}
+2a_{12}\Big<n_{i}\partial_{\rho}R^{(1)}\partial^{\rho}\partial_{0}\partial_{q}\tilde{h}_{iq}^{TT}\Big>^{[2]}\bigg),\\
\label{equ4.44}\Big<n_{i}(\tau^{ia}_{\rm{rad}})^{(2)}\Big>^{[2]}&=\Big<n_{i}(\tau^{ia}_{\rm{rad}})^{(2)}_{f(R)}\Big>^{[2]}
+\frac{1}{\kappa}\bigg(-\frac{a_{12}}{9a_{11}}\Big<n_{a}{\left(R^{(1)}\right)}^2\Big>^{[2]}
+4a_{11}a_{12}\Big<n_{a}R^{(1)}\square_{\eta}^{2}R^{(1)}\Big>^{[2]}\notag\\
&+\frac{4a_{12}}{3}\Big<n_{i}R^{(1)}\partial_{i}\partial_{a}R^{(1)}\Big>^{[2]}
-8a_{11}a_{12}\Big<n_{i}R^{(1)}\partial_{i}\partial_{a}\square_{\eta}R^{(1)}\Big>^{[2]}
+2a_{12}\Big<n_{i}\partial_{p}R^{(1)}\partial_{i}\partial_{a}\partial_{q}\tilde{h}_{pq}^{TT}\Big>^{[2]}\notag\\
&+2a_{12}\Big<n_{i}\partial_{\sigma}R^{(1)}\partial^{\sigma}\square_{\eta}\tilde{h}_{ia}^{TT}\Big>^{[2]}
-2a_{12}\Big<n_{i}\partial_{\rho}R^{(1)}\partial^{\rho}\partial_{a}\partial_{q}\tilde{h}_{iq}^{TT}\Big>^{[2]}
-2a_{12}\Big<n_{i}\partial_{\sigma}R^{(1)}\partial^{\sigma}\partial_{i}\partial_{q}\tilde{h}_{aq}^{TT}\Big>^{[2]}\bigg).
\end{align}
From Eq.~\eqref{equ3.14}, the de Donder condition under the TT gauge is
\begin{align}
\label{equ4.45}\partial_{p}\tilde{h}_{pq}^{TT}=0,
\end{align}
and then, with the help of the Eq.~\eqref{equ3.58} in the radiation field and Eq.~\eqref{equ3.60}, Eqs.~\eqref{equ4.43} and~\eqref{equ4.44} reduce to
\begin{align}
\label{equ4.46}\Big<n_{i}(\tau^{i0}_{\rm{rad}})^{(2)}\Big>^{[2]}&=\Big<n_{i}(\tau^{i0}_{\rm{rad}})^{(2)}_{f(R)}\Big>^{[2]},\\
\label{equ4.47}\Big<n_{i}(\tau^{ia}_{\rm{rad}})^{(2)}\Big>^{[2]}&=\Big<n_{i}(\tau^{ia}_{\rm{rad}})^{(2)}_{f(R)}\Big>^{[2]}.
\end{align}
Inserting them into Eqs.~\eqref{equ4.30} and~\eqref{equ4.31}, respectively, gives rise to
\begin{align}
\label{equ4.48}\left(\frac{dE}{d\Omega dt}\right)^{[0]}&=cr^2\Big<n_{i}(\tau^{i0}_{\rm{rad}})^{(2)}_{f(R)}\Big>^{[2]}=\left(\frac{dE}{d\Omega dt}\right)^{[0]}_{f(R)},\\
\label{equ4.49}\left(\frac{dP_{a}}{d\Omega dt}\right)^{[0]}&=r^2\Big<n_{i}(\tau^{ia}_{\rm{rad}})^{(2)}_{f(R)}\Big>^{[2]}=\left(\frac{dP_{a}}{d\Omega dt}\right)^{[0]}_{f(R)},
\end{align}
where $\left(\frac{dE}{d\Omega dt}\right)^{[0]}_{f(R)}$ and $\left(\frac{dP_{a}}{d\Omega dt}\right)^{[0]}_{f(R)}$ are
the fluxes of energy and momentum carried by the outward-propagating GW in linearized $f(R)$ gravity, respectively, and thus, according to the Ref. II~\cite{Wu:2018hjx},
\begin{align}
\label{equ4.50}\left(\frac{dE}{d\Omega dt}\right)^{[0]}&=\frac{c r^2}{4\kappa}\Big<\partial_{0}\tilde{h}_{pq}^{TT}\partial_{0}\tilde{h}_{pq}^{TT}\Big>^{[2]}-\frac{6cr^2a_{11}^2}{\kappa}\Big<\partial_{r}R^{(1)}\partial_{0}R^{(1)}\Big>^{[2]},\\
\label{equ4.51}\left(\frac{dP_{a}}{d\Omega dt}\right)^{[0]}&=-\frac{ r^2}{4\kappa}\Big<\partial_{0}\tilde{h}_{pq}^{TT}\partial_{a}\tilde{h}_{pq}^{TT}\Big>^{[2]}+\frac{6r^2a_{11}^2}{\kappa}\Big<\partial_{r}R^{(1)}\partial_{a}R^{(1)}\Big>^{[2]}.
\end{align}
The calculation of the average part in Eq.~\eqref{equ4.36} is similar to the above process.
\begin{align}
\label{equ4.52}\Big<\epsilon_{abc}n_{i}n_{b}(\tau^{ic}_{\rm{rad}})^{(2)}\Big>^{[2]}=&\Big<\epsilon_{abc}n_{i}n_{b}(\tau^{ic}_{\rm{rad}})^{(2)}_{f(R)}\Big>^{[2]}
+\frac{1}{\kappa}\bigg(
\frac{4a_{12}}{3}\Big<\epsilon_{abc}n_{i}n_{b}R^{(1)}\partial_{i}\partial_{c}R^{(1)}\Big>^{[2]}\notag\\
&-8a_{11}a_{12}\Big<\epsilon_{abc}n_{i}n_{b}\partial_{i}\partial_{\alpha}R^{(1)}\partial_{c}\partial^{\alpha}R^{(1)}\Big>^{[2]}
-2a_{12}\Big<\epsilon_{abc}n_{i}n_{b}\partial_{p}\partial_{q}R^{(1)}\partial_{i}\partial_{c}\tilde{h}_{pq}^{TT}\Big>^{[2]}\notag\\
&-2a_{12}\Big<\epsilon_{abc}n_{i}n_{b}\partial_{\rho}\partial_{\sigma}R^{(1)}\partial^{\rho}\partial^{\sigma}\tilde{h}_{ic}^{TT}\Big>^{[2]}
+2a_{12}\Big<\epsilon_{abc}n_{i}n_{b}\partial_{\rho}\partial_{q}R^{(1)}\partial^{\rho}\partial_{c}\tilde{h}_{iq}^{TT}\Big>^{[2]}\notag\\
&+2a_{12}\Big<\epsilon_{abc}n_{i}n_{b}\partial_{\sigma}\partial_{q}R^{(1)}\partial^{\sigma}\partial_{i}\tilde{h}_{cq}^{TT}\Big>^{[2]}\bigg)\notag\\
=&\Big<\epsilon_{abc}n_{i}n_{b}(\tau^{ic}_{\rm{rad}})^{(2)}_{f(R)}\Big>^{[2]}
+\frac{1}{\kappa}\bigg(
\frac{4a_{12}}{3}\Big<\epsilon_{abc}n_{i}n_{b}R^{(1)}\partial_{i}\partial_{c}R^{(1)}\Big>^{[2]}\notag\\
&-8a_{11}a_{12}\Big<\epsilon_{abc}n_{i}n_{b}R^{(1)}\partial_{i}\partial_{c}\square_{\eta}R^{(1)}\Big>^{[2]}
+2a_{12}\Big<\epsilon_{abc}n_{i}n_{b}\partial_{p}R^{(1)}\partial_{i}\partial_{c}\partial_{q}\tilde{h}_{pq}^{TT}\Big>^{[2]}\notag\\
&+2a_{12}\Big<\epsilon_{abc}n_{i}n_{b}\partial_{\sigma}R^{(1)}\partial^{\sigma}\square_{\eta}\tilde{h}_{ic}^{TT}\Big>^{[2]}
-2a_{12}\Big<\epsilon_{abc}n_{i}n_{b}\partial_{\rho}R^{(1)}\partial^{\rho}\partial_{c}\partial_{q}\tilde{h}_{iq}^{TT}\Big>^{[2]}\notag\\
&-2a_{12}\Big<\epsilon_{abc}n_{i}n_{b}\partial_{\sigma}R^{(1)}\partial^{\sigma}\partial_{i}\partial_{q}\tilde{h}_{cq}^{TT}\Big>^{[2]}\bigg)\notag\\
=&\Big<\epsilon_{abc}n_{i}n_{b}(\tau^{ic}_{\rm{rad}})^{(2)}_{f(R)}\Big>^{[2]},
\end{align}
and then, by the result in Ref. II~\cite{Wu:2018hjx}, Eq.~\eqref{equ4.36} holds, which ensures that the flux of angular momentum carried by the outward-propagating GW does not diverge.

Now, we begin to calculate the average part in Eq.~(\ref{equ4.32}). Eq.~(\ref{equ4.38}) results in
\begin{align}
\label{equ4.53}\Big<\epsilon_{abc}n_{i}n_{b}(\tau^{ic}_{\rm{rad}})^{(2)}\Big>^{[3]}=&\Big<\epsilon_{abc}n_{i}n_{b}(\tau^{ic}_{\rm{rad}})^{(2)}_{f(R)}\Big>^{[3]}
+\frac{1}{\kappa}\bigg(
\frac{4a_{12}}{3}\Big<\epsilon_{abc}n_{i}n_{b}R^{(1)}\partial_{i}\partial_{c}R^{(1)}\Big>^{[3]}\notag\\
&-8a_{11}a_{12}\Big<\epsilon_{abc}n_{i}n_{b}\partial_{i}\partial_{\alpha}R^{(1)}\partial_{c}\partial^{\alpha}R^{(1)}\Big>^{[3]}
-2a_{12}\Big<\epsilon_{abc}n_{i}n_{b}\partial_{p}\partial_{q}R^{(1)}\partial_{i}\partial_{c}\tilde{h}_{pq}^{TT}\Big>^{[3]}\notag\\
&-2a_{12}\Big<\epsilon_{abc}n_{i}n_{b}\partial_{\rho}\partial_{\sigma}R^{(1)}\partial^{\rho}\partial^{\sigma}\tilde{h}_{ic}^{TT}\Big>^{[3]}
+2a_{12}\Big<\epsilon_{abc}n_{i}n_{b}\partial_{\rho}\partial_{q}R^{(1)}\partial^{\rho}\partial_{c}\tilde{h}_{iq}^{TT}\Big>^{[3]}\notag\\
&+2a_{12}\Big<\epsilon_{abc}n_{i}n_{b}\partial_{\sigma}\partial_{q}R^{(1)}\partial^{\sigma}\partial_{i}\tilde{h}_{cq}^{TT}\Big>^{[3]}\bigg).
\end{align}
By Eqs.~\eqref{equ3.66} and \eqref{equ3.67}, the average part of the third term in Eq.~\eqref{equ4.53} is
\begin{align}
\label{equ4.54}\Big<\epsilon_{abc}n_{i}n_{b}\partial_{i}\partial_{\alpha}R^{(1)}\partial_{c}\partial^{\alpha}R^{(1)}\Big>^{[3]}=&
-\Big<\epsilon_{abc}n_{i}n_{b}\partial_{\alpha}R^{(1)}\partial_{i}\partial_{c}\partial^{\alpha}R^{(1)}\Big>^{[3]}
-\Big<\epsilon_{abc}(\partial_{i}n_{i})n_{b}\partial_{\alpha}R^{(1)}\partial_{c}\partial^{\alpha}R^{(1)}\Big>^{[3]}\notag\\
&-\Big<\epsilon_{abc}n_{i}(\partial_{i}n_{b})\partial_{\alpha}R^{(1)}\partial_{c}\partial^{\alpha}R^{(1)}\Big>^{[3]}\notag\\
=&\Big<\epsilon_{abc}n_{i}n_{b}R^{(1)}\partial_{i}\partial_{c}\square_{\eta}R^{(1)}\Big>^{[3]}
+\Big<\epsilon_{abc}(\partial_{p}n_{i})n_{b}R^{(1)}\partial_{i}\partial_{c}\partial_{p}R^{(1)}\Big>^{[3]}\notag\\
&+\Big<\epsilon_{abc}n_{i}(\partial_{p}n_{b})R^{(1)}\partial_{i}\partial_{c}\partial_{p}R^{(1)}\Big>^{[3]}
+\frac{2}{r}\Big<\epsilon_{abc}n_{b}R^{(1)}\partial_{c}\square_{\eta}R^{(1)}\Big>^{[2]}\notag\\
=&\frac{1}{6a_{11}}\Big<\epsilon_{abc}n_{i}n_{b}R^{(1)}\partial_{i}\partial_{c}R^{(1)}\Big>^{[3]}
+\frac{1}{r}\Big<\epsilon_{abc}n_{b}R^{(1)}\partial_{c}\Delta R^{(1)}\Big>^{[2]}\notag\\
&-\frac{2}{r}\Big<\epsilon_{abc}n_{p}n_{i}n_{b}R^{(1)}\partial_{i}\partial_{c}\partial_{p}R^{(1)}\Big>^{[2]}
+\frac{1}{3a_{11}r}\Big<\epsilon_{abc}n_{b}R^{(1)}\partial_{c}R^{(1)}\Big>^{[2]},
\end{align}
where $\Delta=\partial_{i}\partial_{i}$, Eq.~(\ref{equ2.13}) has been used in the second step, and Eqs.~(\ref{equ3.60}) and $\epsilon_{abc}\partial_{b}\partial_{c}=0$ have been used in the third step. Equations (\ref{equ2.13}) and \eqref{equ4.18} bring about
\begin{align}
\label{equ4.55}\left(\partial_{k}R^{(1)}\right)^{[1]}&=\left(\partial_{k}(R^{(1)})^{[1]}\right)^{[1]}=n_{k}\left(\partial_{r}(R^{(1)})^{[1]}\right)^{[1]},\\
\label{equ4.56}\left(\partial_{j}\partial_{k}R^{(1)}\right)^{[1]}&=\left(\partial_{j}(\partial_{k}R^{(1)})^{[1]}\right)^{[1]}=
n_{k}\left(\partial_{j}\left(\partial_{r}(R^{(1)})^{[1]}\right)^{[1]}\right)^{[1]}=n_{j}n_{k}\left(\partial_{r}\left(\partial_{r}(R^{(1)})^{[1]}\right)^{[1]}\right)^{[1]},\\
\label{equ4.57}\left(\partial_{i}\partial_{j}\partial_{k}R^{(1)}\right)^{[1]}&=\left(\partial_{i}(\partial_{j}\partial_{k}R^{(1)})^{[1]}\right)^{[1]}=
n_{j}n_{k}\left(\partial_{i}\left(\partial_{r}\left(\partial_{r}(R^{(1)})^{[1]}\right)^{[1]}\right)^{[1]}\right)^{[1]}\notag\\
&=n_{i}n_{j}n_{k}\left(\partial_{r}\left(\partial_{r}\left(\partial_{r}(R^{(1)})^{[1]}\right)^{[1]}\right)^{[1]}\right)^{[1]},
\end{align}
and then, there are
\begin{align}
\label{equ4.58}\Big<\epsilon_{abc}n_{b}R^{(1)}\partial_{c}\Delta R^{(1)}\Big>^{[2]}&=\Big<\epsilon_{abc}n_{b}(R^{(1)})^{[1]}(\partial_{c}\Delta R^{(1)})^{[1]}\Big>^{[2]}=0,\\
\label{equ4.59}\Big<\epsilon_{abc}n_{p}n_{i}n_{b}R^{(1)}\partial_{i}\partial_{c}\partial_{p}R^{(1)}\Big>^{[2]}&=\Big<\epsilon_{abc}n_{p}n_{i}n_{b}(R^{(1)})^{[1]}(\partial_{i}\partial_{c}\partial_{p}R^{(1)})^{[1]}\Big>^{[2]}=0,\\
\label{equ4.60}\Big<\epsilon_{abc}n_{b}R^{(1)}\partial_{c}R^{(1)}\Big>^{[2]}&=\Big<\epsilon_{abc}n_{b}(R^{(1)})^{[1]}(\partial_{c}R^{(1)})^{[1]}\Big>^{[2]}=0,
\end{align}
where $\epsilon_{abc}n_bn_c=0$ have been used, so by Eq.~\eqref{equ4.54}, the average part of the third term in Eq.~\eqref{equ4.53} is
\begin{align}
\label{equ4.61}\Big<\epsilon_{abc}n_{i}n_{b}\partial_{i}\partial_{\alpha}R^{(1)}\partial_{c}\partial^{\alpha}R^{(1)}\Big>^{[3]}
=\frac{1}{6a_{11}}\Big<\epsilon_{abc}n_{i}n_{b}R^{(1)}\partial_{i}\partial_{c}R^{(1)}\Big>^{[3]}.
\end{align}

By using Eqs.~\eqref{equ3.66} and \eqref{equ3.67} again, the average part of the fourth term in Eq.~\eqref{equ4.53} is
\begin{align}
\label{equ4.62}\Big<\epsilon_{abc}n_{i}n_{b}\partial_{p}\partial_{q}R^{(1)}\partial_{i}\partial_{c}\tilde{h}_{pq}^{TT}\Big>^{[3]}=&
-\Big<\epsilon_{abc}(\partial_{p}n_{i})n_{b}\partial_{q}R^{(1)}\partial_{i}\partial_{c}\tilde{h}_{pq}^{TT}\Big>^{[3]}
-\Big<\epsilon_{abc}n_{i}(\partial_{p}n_{b})\partial_{q}R^{(1)}\partial_{i}\partial_{c}\tilde{h}_{pq}^{TT}\Big>^{[3]}\notag\\
=&\frac{2}{r}\Big<\epsilon_{abc}n_{p}n_{i}n_{b}\partial_{q}R^{(1)}\partial_{i}\partial_{c}\tilde{h}_{pq}^{TT}\Big>^{[2]}
-\frac{1}{r}\Big<\epsilon_{abc}n_{i}\partial_{q}R^{(1)}\partial_{i}\partial_{c}\tilde{h}_{bq}^{TT}\Big>^{[2]}\notag\\
=&-\frac{2}{r}\Big<\epsilon_{abc}n_{p}n_{i}n_{b}R^{(1)}\partial_{i}\partial_{c}\partial_{q}\tilde{h}_{pq}^{TT}\Big>^{[2]}
+\frac{1}{r}\Big<\epsilon_{abc}n_{i}R^{(1)}\partial_{i}\partial_{c}\partial_{q}\tilde{h}_{bq}^{TT}\Big>^{[2]}=0,
\end{align}
where Eqs.~\eqref{equ2.13} and~\eqref{equ4.45} have been used. Similar reason can be used to calculate average parts of the fifth, the sixth, and the seventh terms in Eq.~\eqref{equ4.53}, namely,
\begin{align}
\label{equ4.63}\Big<\epsilon_{abc}n_{i}n_{b}\partial_{\rho}\partial_{\sigma}R^{(1)}\partial^{\rho}\partial^{\sigma}\tilde{h}_{ic}^{TT}\Big>^{[3]}
=&-\Big<\epsilon_{abc}(\partial_{\rho}n_{i})n_{b}\partial_{\sigma}R^{(1)}\partial^{\rho}\partial^{\sigma}\tilde{h}_{ic}^{TT}\Big>^{[3]}
-\Big<\epsilon_{abc}n_{i}(\partial_{\rho}n_{b})\partial_{\sigma}R^{(1)}\partial^{\rho}\partial^{\sigma}\tilde{h}_{ic}^{TT}\Big>^{[3]}\notag\\
=&-\Big<\epsilon_{abc}(\partial_{q}n_{i})n_{b}\partial_{\sigma}R^{(1)}\partial_{q}\partial^{\sigma}\tilde{h}_{ic}^{TT}\Big>^{[3]}
-\Big<\epsilon_{abc}n_{i}(\partial_{q}n_{b})\partial_{\sigma}R^{(1)}\partial_{q}\partial^{\sigma}\tilde{h}_{ic}^{TT}\Big>^{[3]}\notag\\
=&\frac{2}{r}\Big<\epsilon_{abc}n_{q}n_{i}n_{b}\partial_{\sigma}R^{(1)}\partial_{q}\partial^{\sigma}\tilde{h}_{ic}^{TT}\Big>^{[2]}
-\frac{1}{r}\Big<\epsilon_{abc}n_{i}\partial_{\sigma}R^{(1)}\partial_{b}\partial^{\sigma}\tilde{h}_{ic}^{TT}\Big>^{[2]}\notag\\
=&-\frac{2}{r}\Big<\epsilon_{abc}n_{q}n_{i}n_{b}R^{(1)}\partial_{q}\square_{\eta}\tilde{h}_{ic}^{TT}\Big>^{[2]}
+\frac{1}{r}\Big<\epsilon_{abc}n_{i}R^{(1)}\partial_{b}\square_{\eta}\tilde{h}_{ic}^{TT}\Big>^{[2]}=0,\\
\label{equ4.64}\Big<\epsilon_{abc}n_{i}n_{b}\partial_{\rho}\partial_{q}R^{(1)}\partial^{\rho}\partial_{c}\tilde{h}_{iq}^{TT}\Big>^{[3]}
=&-\Big<\epsilon_{abc}(\partial_{q}n_{i})n_{b}\partial_{\rho}R^{(1)}\partial^{\rho}\partial_{c}\tilde{h}_{iq}^{TT}\Big>^{[3]}
-\Big<\epsilon_{abc}n_{i}(\partial_{q}n_{b})\partial_{\rho}R^{(1)}\partial^{\rho}\partial_{c}\tilde{h}_{iq}^{TT}\Big>^{[3]}\notag\\
=&\frac{2}{r}\Big<\epsilon_{abc}n_{q}n_{i}n_{b}\partial_{\rho}R^{(1)}\partial^{\rho}\partial_{c}\tilde{h}_{iq}^{TT}\Big>^{[2]}
-\frac{1}{r}\Big<\epsilon_{abc}n_{i}\partial_{\rho}R^{(1)}\partial^{\rho}\partial_{c}\tilde{h}_{ib}^{TT}\Big>^{[2]}\notag\\
=&-\frac{2}{r}\Big<\epsilon_{abc}n_{q}n_{i}n_{b}R^{(1)}\partial_{c}\square_{\eta}\tilde{h}_{iq}^{TT}\Big>^{[2]}
+\frac{1}{r}\Big<\epsilon_{abc}n_{i}R^{(1)}\partial_{c}\square_{\eta}\tilde{h}_{ib}^{TT}\Big>^{[2]}=0,\\
\label{equ4.65}\Big<\epsilon_{abc}n_{i}n_{b}\partial_{\sigma}\partial_{q}R^{(1)}\partial^{\sigma}\partial_{i}\tilde{h}_{cq}^{TT}\Big>^{[3]}
=&-\Big<\epsilon_{abc}(\partial_{q}n_{i})n_{b}\partial_{\sigma}R^{(1)}\partial^{\sigma}\partial_{i}\tilde{h}_{cq}^{TT}\Big>^{[3]}
-\Big<\epsilon_{abc}n_{i}(\partial_{q}n_{b})\partial_{\sigma}R^{(1)}\partial^{\sigma}\partial_{i}\tilde{h}_{cq}^{TT}\Big>^{[3]}\notag\\
=&\frac{2}{r}\Big<\epsilon_{abc}n_{q}n_{i}n_{b}\partial_{\sigma}R^{(1)}\partial^{\sigma}\partial_{i}\tilde{h}_{cq}^{TT}\Big>^{[2]}
=-\frac{2}{r}\Big<\epsilon_{abc}n_{q}n_{i}n_{b}R^{(1)}\partial_{i}\square_{\eta}\tilde{h}_{cq}^{TT}\Big>^{[2]}=0,
\end{align}
where Eq.~\eqref{equ3.58} in the radiation field has been used in the above derivations, and $\tilde{h}_{ii}^{TT}=0$ and $\epsilon_{abc}\tilde{h}_{bc}^{TT}=0$ have been used in the derivations of Eqs.~\eqref{equ4.64} and~\eqref{equ4.65}, respectively.
Therefore, Eq.~\eqref{equ4.53} reduces to
\begin{align}
\label{equ4.66}\Big<\epsilon_{abc}n_{i}n_{b}(\tau^{ic}_{\rm{rad}})^{(2)}\Big>^{[3]}=&\Big<\epsilon_{abc}n_{i}n_{b}(\tau^{ic}_{\rm{rad}})^{(2)}_{f(R)}\Big>^{[3]}.
\end{align}
Inserting it into Eq.~\eqref{equ4.32} gives rise to
\begin{align}
\label{equ4.67}\left(\frac{dJ_{a}}{d\Omega dt}\right)^{[0]}&=r^3\Big<\epsilon_{abc}n_{i}n_{b}(\tau^{ic}_{\rm{rad}})^{(2)}_{f(R)}\Big>^{[3]}=\left(\frac{dJ_{a}}{d\Omega dt}\right)^{[0]}_{f(R)},
\end{align}
where $\left(\frac{dJ_{a}}{d\Omega dt}\right)^{[0]}_{f(R)}$ is
the flux of angular momentum carried by the outward-propagating GW in linearized $f(R)$ gravity, and thus, according to the Ref. II~\cite{Wu:2018hjx},
\begin{align}
\label{equ4.68}\left(\frac{dJ_{a}}{d\Omega dt}\right)^{[0]}&=\frac{r^2}{2\kappa}\Big<\epsilon_{abc}\Big(\tilde{h}_{bi}^{TT}\partial_{0}\tilde{h}_{ic}^{TT}-\frac{1}{2}x_{b}\partial_{c}\tilde{h}_{pq}^{TT}\partial_{0}\tilde{h}_{pq}^{TT}\Big)\Big>^{[2]}
+\frac{6r^3a_{11}^2}{\kappa}\Big<\epsilon_{abc}n_{b}\partial_{r}R^{(1)}\partial_{c}R^{(1)}\Big>^{[3]}.
\end{align}

The coupling constant $a_{12}$ in the Lagrangian \eqref{equ1.1} appears in the expressions of $(\tau^{i0}_{\rm{rad}})^{(2)}$ and $(\tau^{ij}_{\rm{rad}})^{(2)}$, which shows that the fluxes of the energy, momentum, and angular momentum carried by GWs in linearized $f(R,\mathcal{G})$ gravity depend on the GB curvature scalar $\mathcal{G}$. However, the above derivations of Eqs.~\eqref{equ4.48}, \eqref{equ4.49}, and \eqref{equ4.67} show that, similarly to that of the effective stress-energy tensor of GWs, the terms associated with the coupling constant $a_{12}$ in the fluxes of the energy, momentum, and angular momentum are cancelled each other after the average over a small spatial volume, and the dependence on $a_{12}$ also vanishes. Therefore, the fluxes of the energy, momentum, and angular momentum carried by GWs in linearized $f(R,\mathcal{G})$ gravity are the same as those in linearized $f(R)$ gravity. Moreover, the fluxes of energy and momentum carried by GW in linearized $f(R,\mathcal{G})$ gravity could also be directly expressed in terms of the effective stress-energy tensor $t^{\mu\nu}$, namely,
\begin{align}
\label{equ4.69}\frac{dE}{d\Omega dt}&=+c r^2n_{i}t^{i0},\qquad
\frac{dP_{a}}{d\Omega dt}=+r^2n_{i}t^{ia}.
\end{align}
Eq.~\eqref{equ3.69} shows that the effective stress-energy tensor of GWs in linearized $f(R,\mathcal{G})$ gravity is the same as that in linearized $f(R)$ gravity, so there are
\begin{align}
\label{equ4.70}\frac{dE}{d\Omega dt}&=+c r^2n_{i}t^{i0}_{f(R)},\qquad
\frac{dP_{a}}{d\Omega dt}=+r^2n_{i}t^{ia}_{f(R)},
\end{align}
and then, according to the Ref. II~\cite{Wu:2018hjx},
both of the leading terms in Eq.~\eqref{equ4.69} can recover Eqs.~(\ref{equ4.50}) and (\ref{equ4.51}), respectively.

Integrating the fluxes of the energy, momentum, and angular momentum carried by GWs in linearized $f(R,\mathcal{G})$ gravity over a sphere can give rise to the total power and rates of momentum and angular momentum, respectively, and then, according to the Ref. II~\cite{Wu:2018hjx} again, they are
\begin{align}
\label{equ4.71}\left(\frac{dE}{dt}\right)^{[0]}&=\left(\frac{dE}{dt}\right)^{[0]}_{GR}-\sum_{l=0}^{\infty}\frac{G}{3c^{2l+4}}\frac{1}{l!(2l+1)!!}\Big<\partial_{t}\hat{K}_{I_{l}}(t,r)\partial_{r}\hat{K}_{I_{l}}(t,r)\Big>,\\
\label{equ4.72}\left(\frac{dP_{a}}{dt}\right)^{[0]}&=\left(\frac{dP_{a}}{dt}\right)^{[0]}_{GR}+\sum_{l=0}^{\infty}\frac{2G}{3c^{2l+5}}\frac{1}{l!(2l+3)!!}\Big<\partial_{r}\hat{K}_{I_{l}}(t,r)\partial_{r}\hat{K}_{aI_{l}}(t,r)\Big>,\\
\label{equ4.73}\left(\frac{dJ_{a}}{dt}\right)^{[0]}&=\left(\frac{dJ_{a}}{dt}\right)^{[0]}_{GR}-\sum_{l=1}^{\infty}\frac{G}{3c^{2l+4}}\frac{1}{(l-1)!(2l+1)!!}\Big<\epsilon_{abc}\hat{K}_{bI_{l-1}}(t,r)\partial_{r}\hat{K}_{cI_{l-1}}(t,r)\Big>,
\end{align}
where
\begin{align}
\label{equ4.74}\bigg(\frac{dE}{dt}\bigg)^{[0]}_{GR}=&\sum_{l=2}^{\infty}\frac{G}{c^{2l+1}}\bigg(
\frac{(l+1)(l+2)}{l(l-1)l!(2l+1)!!}\Big<\partial_{u}^{l+1}\hat{M}_{I_{l}}(u)\partial_{u}^{l+1}\hat{M}_{I_{l}}(u)\Big>\notag\\
&\qquad\qquad\qquad\qquad\qquad+\frac{4l(l+2)}{c^2(l-1)(l+1)!(2l+1)!!}\Big<\partial_{u}^{l+1}\hat{S}_{I_{l}}(u)\partial_{u}^{l+1}\hat{S}_{I_{l}}(u)\Big>\bigg),
\end{align}
\begin{align}
\label{equ4.75}\bigg(\frac{dP_{a}}{dt}\bigg)^{[0]}_{GR}=&\sum_{l=2}^{\infty}\frac{2G}{c^{2l+3}}\bigg(
\frac{(l+2)(l+3)}{l(l+1)!(2l+3)!!}\Big<\partial_{u}^{l+1}\hat{M}_{I_{l}}(u)\partial_{u}^{l+2}\hat{M}_{aI_{l}}(u)\Big>\notag\\
&\qquad\qquad\qquad\qquad\qquad+\frac{4(l+2)}{(l-1)(l+1)!(2l+1)!!}\Big<\epsilon_{abc}\partial_{u}^{l+1}\hat{M}_{bI_{l-1}}(u)\partial_{u}^{l+1}\hat{S}_{cI_{l-1}}(u)\Big>\notag\\
&\qquad\qquad\qquad\qquad\qquad+\frac{4(l+3)}{c^2(l+1)!(2l+3)!!}\Big<\partial_{u}^{l+1}\hat{S}_{I_{l}}(u)\partial_{u}^{l+2}\hat{S}_{aI_{l}}(u)\Big>\bigg),\\
\label{equ4.76}\bigg(\frac{dJ_{a}}{dt}\bigg)^{[0]}_{GR}=&\sum_{l=2}^{\infty}\frac{G}{c^{2l+1}}\bigg(
\frac{(l+1)(l+2)}{(l-1)l!(2l+1)!!}\Big<\epsilon_{abc}\partial_{u}^{l}\hat{M}_{bI_{l-1}}(u)\partial_{u}^{l+1}\hat{M}_{cI_{l-1}}(u)\Big>\notag\\
&\qquad\qquad\qquad\qquad\qquad+\frac{4l^2(l+2)}{c^2(l-1)(l+1)!(2l+1)!!}\Big<\epsilon_{abc}\partial_{u}^{l}\hat{S}_{bI_{l-1}}(u)\partial_{u}^{l+1}\hat{S}_{cI_{l-1}}(u)\Big>\bigg)
\end{align}
are the corresponding results in GR with the replacement of the variables $h_{pq}^{TT}$ by $\tilde{h}_{pq}^{TT}$.

Obviously, these expressions include two parts, which are associated with the tensor part $\tilde{h}^{\mu\nu}$ and the scalar part $R^{(1)}$ in the multipole expansion of linearized $f(R,\mathcal{G})$ gravity, respectively. The former are the GR-like part, and the latter, contributed by the $R^2$ term in the Lagrangian~\eqref{equ1.1}, are the corrections to the results in GR, which implies that the GB curvature scalar $\mathcal{G}$ does not contribute to the energy, momentum, and angular momentum carried by GWs in linearized $f(R,\mathcal{G})$ gravity.

\section{Conclusions and discussions \label{Sec:Conclusion}}
In this paper, similarly to GR~\cite{Thorne:1980ru,Blanchet:2013haa} and $f(R)$ gravity~\cite{Wu:2017vvm},
the field equations of $f(R,\mathcal{G})$ gravity have been rewritten in the form of obvious wave equations in a fictitious flat spacetime under the de Donder condition, even though the effective gravitational field amplitude $\tilde h^{\mu\nu}$ is not a perturbation.  The corresponding source term of this wave equation, composed of all the nonlinear terms under the post-Minkowskian method, is obtained, and it is the stress-energy pseudotensor of the matter fields and the gravitational field. For the linearized $f(R,\mathcal{G})$ gravity, the corresponding field equations, as the linear results, are the same as those of the linearized $f(R)$ gravity, as is expected. Under the post-Minkowskian method, the coupling constant $a_{12}$ in the Lagrangian \eqref{equ1.1} appears in the second-order field equations of $f(R,\mathcal{G})$ gravity, which implies that the GB curvature scalar $\mathcal{G}$ usually plays an important role in the nonlinear effects. Thus, as a typical second-order nonlinear quantity, the effective stress-energy tensor of GWs in linearized $f(R,\mathcal{G})$ gravity should include the contribution from $\mathcal{G}$. However,
it is shown that $\mathcal{G}$ has nothing to do with the effective stress-energy tensor of GWs in linearized $f(R,\mathcal{G})$ gravity, and the effective stress-energy tensor of GWs in linearized $f(R,\mathcal{G})$ gravity is the same as that in linearized $f(R)$ gravity.

Besides the effective stress-energy tensor of GWs, the energy, momentum, and angular momentum carried by GWs in linearized $f(R,\mathcal{G})$ gravity, as the further second-order nonlinear quantities, need to be discussed. As in
GR~\cite{Thorne:1980ru} and $f(R)$ gravity~\cite{Wu:2018hjx}, the above effective stress-energy tensor of GWs can be used to evaluate the fluxes of the energy and the momentum but not to evaluate the flux of the angular momentum~\cite{Thorne:1980ru}. Therefore, we need to find a new way, not depending on the effective stress-energy tensor of GWs, to deal with the angular momentum. According to the Refs.~\cite{Thorne:1980ru,Peters:1964zz,Wu:2018hjx}, we can deal with the energy, momentum, and angular momentum carried by GWs in linearized $f(R,\mathcal{G})$ gravity in a unified way. What this way requires is the stress-energy pseudotensor of $f(R,\mathcal{G})$ gravity in the radiation field, instead of the effective stress-energy tensor of GWs. By the above unified way, the multipole expansion of the linearized $f(R,\mathcal{G})$ gravity need to be discussed firstly. The definitions of the gravitational field amplitude $h^{\mu\nu}$ and the effective gravitational field amplitude $\tilde{h}^{\mu\nu}$ show that they satisfy the same relation, and especially, the linearized one~\eqref{equ1.4} as that in $f(R)$ gravity. Further, for the linearized $f(R,\mathcal{G})$ gravity and the linearized $f(R)$ gravity, $\tilde{h}^{\mu\nu}$ and $R^{(1)}$ satisfy the same field equations, respectively, and
therefore, these two gravitational models have the same multipole expansion, and in particular, they have the same multipole expansion in the radiation field.

With the help of the relevant STF technique, the $1/r$ expansion in the distance to the source can be applied to the linearized $f(R,\mathcal{G})$ gravity. Based on this, the energy, momentum, and angular momentum carried by GWs in linearized $f(R,\mathcal{G})$ gravity can be dealt with. In the expressions of the fluxes of the energy, momentum, and angular momentum carried by GWs in linearized $f(R,\mathcal{G})$ gravity, the existence of the coupling constant $a_{12}$ in the Lagrangian \eqref{equ1.1} implies the GB curvature scalar $\mathcal{G}$ should play an important role.
However, because of the average over a small spatial volume, the terms associated with the coupling constant $a_{12}$ in these expressions are cancelled each other, and the dependence on $a_{12}$  vanishes. Therefore, $\mathcal{G}$ does not contribute to the energy, momentum, and angular momentum carried by GWs in linearized $f(R,\mathcal{G})$ gravity,
which implies that these fluxes are the same as those in linearized $f(R)$ gravity. Thus, by integrating these fluxes over a sphere, we present the general expressions of the total power and rates of momentum and angular momentum in the GWs for linearized $f(R,\mathcal{G})$ gravity. Same as those in $f(R)$ gravity, these expressions include the GR-like parts and the corrections, contributed by the $R^2$ term in the Lagrangian~\eqref{equ1.1}, to the corresponding results in GR.

As mentioned above, in terms of the energy, momentum, and angular momentum carried by GWs, the linearized $f(R,\mathcal{G})$ gravity behave as the linearized $f(R)$ gravity, but for $f(R,\mathcal{G})$ gravity, where is the contribution of the GB curvature scalar $\mathcal{G}$? One obvious answer is that $\mathcal{G}$ perhaps plays an important role in the higher-order nonlinear effects. The similar situation also appears for the $f(R,\mathcal{G})$ gravity under the weak-field and slow-motion method~\cite{Wu:2015maa}. Under this case, the GB curvature $\mathcal{G}$ has nothing to do with the effects at $(v/c)^{2}$ and $(v/c)^{3}$ orders, and only contributes to the effects at $(v/c)^{4}$ order or the higher-order, where $v/c$ is the perturbation parameter of the weak-field and slow-motion method. In the present gravitational Lagrangian~\eqref{equ1.1}, the GB curvature scalar $\mathcal{G}$ works only by the term $a_{12}R\mathcal{G}$, so $f(R,\mathcal{G})$ in Eq.~\eqref{equ1.1} is equivalent to $f(R)+a_{12}R\mathcal{G}$.
Further, $f(R)$ gravity in the metric formalism can be cast in the form of Brans-Dicke (BD) theory with a potential for the effective scalar-field degree of freedom (scalaron)~\cite{DeFelice:2010aj}, and thus, the
theory with Lagrangian $f(R)+a_{12}R\mathcal{G}$ can be cast as a large class of GB gravity whose Lagrangian is equivalent to above BD theory coupled by the GB curvature scalar $\mathcal{G}$. Therefore, the conclusion in this paper also holds for such GB gravity.

\acknowledgments{This work is supported by the National Natural Science Foundation of China (Grant No.~11690022) and by the Strategic Priority Research Program of the Chinese Academy of Sciences ``Multi-waveband Gravitational Wave
Universe" (Grant No. XDB23040000).}

\end{document}